\documentclass[aps,showpacs,amsmath,amssymb,pre,superscriptaddress,twocolumn]{revtex4}
\usepackage{graphicx}
\usepackage{epsfig}
\usepackage{amsmath}
\usepackage{amssymb}

\begin{document}

\title{Tight-binding parameters for charge transfer along DNA}

\author{L.G.D.~Hawke}
\affiliation{Materials Science Department, University of Patras,
Rio GR-26504, Greece}

\author{G.~Kalosakas}
\affiliation{Materials Science Department, University of Patras,
Rio GR-26504, Greece}

\author{C.~Simserides}
\affiliation{Institute of Materials Science,
National Center of Scientific Research Demokritos, GR-15310 Athens, Greece}


\begin{abstract}
We systematically examine all the tight-binding parameters
pertinent to charge transfer along DNA.
The $\pi$ molecular structure of the four DNA bases (adenine, thymine,
cytosine, and guanine) is investigated by using the linear combination of
atomic orbitals method with a recently introduced parametrization.
The HOMO and LUMO wavefunctions and energies
of DNA bases are discussed and then used for calculating
the corresponding wavefunctions of the two B-DNA base-pairs
(adenine-thymine and  guanine-cytosine).
The obtained HOMO and LUMO energies of the bases are in
good agreement with available experimental values.
Our results are then used for estimating the complete set of charge transfer
parameters between neighboring bases and also between successive base-pairs,
considering all possible combinations between them, for both electrons and holes.
The calculated microscopic quantities can be used in mesoscopic theoretical
models of electron or hole transfer along the DNA double helix, as
they provide the necessary parameters for a tight-binding phenomenological
description based on the $\pi$ molecular overlap.
We find that usually the hopping parameters for holes are higher in magnitude
compared to the ones for electrons, which probably indicates that hole transport
along DNA is more favorable than electron transport.
Our findings are also compared with existing calculations from first principles.
\end{abstract}

\maketitle

\section{Introduction}
\label{sec:introduction}
DNA plays a fundamental role in genetics and molecular biology
since its sequence of bases, adenine (A), guanine (G), cytosine (C),
and thymine (T), contains the genetic code of living organisms.
During the last decade, DNA and its charge transport properties
have attracted the interest of a large interdisciplinary community
due to its potential use for nanodevices, either
for assembling nanocircuits or as a molecular wire
\cite{dekker_ratner_2001,keren_et_al_2002,seeman_2003,endres_et_al_2004}.
Charge migration through DNA could also play an important role in biology;
it may be a critical issue in carcinogenesis and mutagenesis
\cite{burrows_muller_1998,cadet_et_al_1994}.
For example the rapid hole migration from other bases to guanine
is connected to the fact that direct strand breaks occur
preferentially at guanines \cite{burrows_muller_1998}.
Long-range charge transfer along the $\pi$-stacking of the DNA double helix
may also be crucial for DNA damage and repair
\cite{dandliker_et_al_1997,rajski_et_al_2000}.

However, it is not yet clear ``how much'' DNA conducts.
Experiments cover a  wide range of behavior.
In particular, previous results found that $\lambda$-DNA covalently bonded
to Au electrodes is an insulator \cite{zhang_et_al_2002}.
Furthermore, insulating behavior
--both for single DNA molecules as well as for small bundles of DNA molecules--
was observed at the 100 nm length scale \cite{storm_et_al_2001};
this was confirmed
both for mixed base-pair sequence as well as for homogeneous poly(dG)-poly(dC),
for lengths between contacts in the range 40-500 nm,
for substrate SiO$_2$ or mica, and for electrode material gold or platinum.
The absence of {\it dc} conductivity in $\lambda$-DNA was also reported for
DNA molecules adsorbed on mica, in agreement with the authors'
first principles electronic structure calculations \cite{de_pablo_et_al_2000}.

In a different direction point the results of Yoo~et~al.~\cite{yoo_et_al_2001},
who investigated the electrical transport through poly(dA)-poly(dT) and
poly(dG)-poly(dC) DNA, i.e. molecules containing identical base-pairs.
Their experimental results suggest that electrical transport through DNA
molecules occurs by polaron hopping and
the possibility of a DNA field-effect transistor
operating at room temperature was also demonstrated \cite{yoo_et_al_2001}.
Moreover, gate-voltage dependent transport measurements showed that
poly(dA)-poly(dT) behaves as a n-type semiconductor, whereas
poly(dG)-poly(dC) behaves as a p-type semiconductor \cite{yoo_et_al_2001}.
In another report it has been shown that poly(dG)-poly(dC) can act as a
semiconducting nanowire exhibiting better conductance than
poly(dA)-poly(dT) \cite{cai_et_al_2000}.

Measurements of electrical current as a function of the potential applied
across single DNA ropes at least 600 nm long, indicated
metallic-like gapless behavior and efficient conduction
comparable to that of conducting polymers \cite{fink_schonenberger_1999}.
In this work it was mentioned that the observed behavior together
with the fact that DNA molecules of specific composition
and length, ranging from a few nucleotides to several tens of $\mu$m,
can be prepared, makes DNA ideally suited for the construction of
mesoscopic electronic devices \cite{fink_schonenberger_1999}.

The {\it ac} conductivity of DNA has been also explored. For example,
$\lambda$-phage DNA at microwave frequencies showed strong temperature
dependence of the {\it ac} conductivity around room temperature, exhibiting a
crossover to a weak temperature dependence at lower
temperatures \cite{tran_et_al_2000}.
Measurements of the quasi-static, frequency-dependent conductivity
below 1 MHz performed on wet-spun, macroscopically oriented,
calf thymus DNA bulk samples, showed that the electrical conductivity
can be rather well described by an activated Arrhenius law with
activation energy of 0.9 eV or by hole hopping \cite{kutnjak_et_al_2003}.

The observed important deviations in the conductivity of DNA can be
--at least partially--
attributed to several external conditions like the type of the substrate,
the distance between the electrodes, and the contact material, in conjunction with
the soft nature of the biomolecule.
In addition, several intrinsic characteristics,
like the local chemical environment which is a result of
the solution involved in the DNA preparation \cite{macnaughton_et_al_2006},
the hydrogen bonding \cite{mallajosyula_et_al_2005}, and the degree of
stretching of DNA \cite{MBKEF_PRB_2002} affect the electronic properties.

The great effort to ameliorate DNA's transport properties
led to alternative molecular conductive candidates based on DNA.
In M-DNA the imino-proton of each base-pair has been substituted
with a metal ion \cite{rakitin_et_al_2001}. In contrast to B-DNA,
M-DNA presents no plateau in its I-V curve and exhibits metallic-like
conduction at room temperature \cite{rakitin_et_al_2001}.
Size-expanded DNA bases are produced by the addition of a benzene ring
to a natural DNA base \cite{fuentes-cabrera_et_al} and have been combined
with natural bases to form xDNA and yDNA, a new class of synthetic nucleic
acids. It has been theoretically shown that xDNA and yDNA
have smaller HOMO-LUMO gap than B-DNA \cite{fuentes-cabrera_et_al},
i.e., they might function as molecular wires better than natural DNA.

Many mechanisms have been suggested to explain the charge transport along
DNA, where generally $\pi$-pathway transfer due to the overlap of
$\pi$ molecular orbitals of the stacked aromatic bases of DNA can
lead to charge propagation even at long distances
\cite{endres_et_al_2004,dandliker_et_al_1997,Barton1996ab}.
Examples of mechanisms that have been examined are band transport
\cite{porath_et_al_2000,adesi_et_al}, polaronic transport
\cite{yoo_et_al_2001,RC,triberis+,KKB,zhang_et_al_2_2002,singh_2004},
fluctuation-facilitated charge migration
\cite{ye_et_al_1999,schuster,bruinsma_et_al_2000,KRB},
and variable range hopping \cite{yu_song_2001}.

The aim of this work is to provide electronic parameters for charge
(electron or hole) transfer along DNA,
assumed that the transport mechanism is the $\pi$-pathway.
Using the linear combination of atomic orbitals (LCAO) method,
our investigation focuses on the calculation of HOMO  (highest occupied
molecular orbital) and LUMO (lowest unoccupied molecular orbital) $\pi$
molecular wavefunctions and energies for the four DNA bases, A, G, C, T,
and the two B-DNA base-pairs, adenine-thymine (A-T) and guanine-cytosine
(G-C). Then the hopping matrix elements between neighboring bases or
base-pairs, for all possible combinations between them, are provided, for both
electrons and holes. The obtained hopping parameters and the HOMO and
LUMO energies can be used in a tight-binding phenomenological
description of charge transfer along DNA. Such a mesoscopic description is applied
in numerous theoretical models of charge transport
\cite{RC,zhang_et_al_2_2002,triberis+,KKB,KRB,yu_song_2001,hennig+,KNF+MKRB,lima}.
Our results are compared with existing calculations from first principles,
as well as with available experimental data for the HOMO and LUMO energies
of DNA bases.

The article is organized in the following way:
In Sec.~\ref{sec:theory} we explain in detail our theoretical approach.
Specifically, in Sec.~\ref{sec:bases} we apply a simple LCAO method for the $\pi$
electronic structure of the four DNA bases using a novel parametrization \cite{HKS},
in Sec.~\ref{sec:base-pairs} we discuss the HOMO and the LUMO of base-pairs
by employing a similar linear combination of molecular orbitals method,
and finally Sec.~\ref{sec:charge-transfer} is devoted to the tight-binding
description and the calculation of charge transfer parameters in
this framework. In Sec.~\ref{sec:resdis} we present and discuss our results,
along with a comparison with earlier theoretical calculations as well as
with available experimental values.
Finally, in Sec.~\ref{sec:conclusions} we state our conclusions.

\section{Theory}
\label{sec:theory}

\subsection{$\pi$ molecular structure of DNA bases}
\label{sec:bases}

In this subsection we present the LCAO methodology used for the description of
the $\pi$ electronic structure of the four isolated DNA bases, A, G, C, and T.
DNA bases are planar organic molecules bonded by $sp^2$
hybridization, where the atoms have their $p_z$ atomic orbitals
perpendicular to the molecular plane.
The electrons that occupy these atomic orbitals are delocalized to form
$\pi$ molecular orbitals. The LCAO method provides a straightforward
approach to obtain the $\pi$ molecular structure.
In its simpler form, used in the present article,
a $\pi$ molecular single-electron wavefunction can be approximated as
\begin{equation}\label{lcao-pi-molecular}
\Psi^{b}({\bf r}) = \sum_{i=1}^N c_i p_z^i ({\bf r}).
\end{equation}
The index $i$ implies summation over all atoms ($N$ totally)
which contribute $p_z$ electrons to the DNA base of interest.
$|c_i|^2$ gives the probability of finding the electron occupying
the molecular orbital $\Psi^{b}({\bf r})$ at the atom indexed by $i$,
while $p_z^i ({\bf r})$ denotes the corresponding atomic, or atomic-like, orbital.
The molecular wavefunction obeys Schr\"{o}dinger equation
$H^{b} \Psi^{b}({\bf r})= E^{b} \Psi^{b}({\bf r})$, where $E^b$ is the
eigenergy of the base.
Substituting in the latter equation $\Psi^{b}({\bf r})$ through
Eq. (\ref{lcao-pi-molecular}), multiplication with
the conjugate orbital $p_z^{j \; \star} ({\bf r})$, and integration
over all space, gives a linear system of $N$ equations
obeyed by the unknown coefficients $c_i$ of the molecular wavefunction
under determination and its energy eigenvalue $E^b$.
Following a standard procedure and  assuming that the atomic or atomic-like
$p_z$ orbitals located at different atoms are orthogonal, one obtains that the
solution of the linear system of equations determining $c_i$ and the corresponding
$E^b$, is equivalent to the diagonalization of the Hamiltonian with matrix elements
$H^b_{ij}=\int d{^3r} \; {p_z^{i\; \star}({\bf r})} \; H^b\; p_z^j ({\bf r})$
(see for example Ref.~\cite{HKS}).

Through numerical diagonalization of the symmetric $N \times N$ Hamiltonian
matrix $H^b_{ij}$ we obtain the coefficients $c_i$ (eigenvectors) providing the
molecular orbitals $\Psi^{b}({\bf r})$ [cf. Eq. (\ref{lcao-pi-molecular})], as well as the
corresponding eigenenergies $E^b$.
To this end we need the values of the Hamiltonian matrix elements,
$H^{b}_{ji}$. We employ a recently introduced parametrization
that has been successfully used for
the description of the energy of $\pi$ frontier orbitals in more than
sixty planar organic molecules \cite{HKS}, including the DNA bases, as
well as in flavin \cite{flavin}, and has shown to provide accurate calculations
of the $\pi$HOMO energies and $\pi$-$\pi^*$ gaps.
Regarding the diagonal matrix elements, $H^{b}_{ii} \equiv \varepsilon_i$,
we use $\varepsilon_C = -6.7$ eV for carbon atoms,
$\varepsilon_{N_2} = -7.9$ eV for nitrogen atoms
contributing one $p_z$ electron (i.e. with coordination number 2),
$\varepsilon_{N_3} = -10.9$ eV for nitrogen atoms
with two $p_z$ electrons (i.e. with coordination number 3),
and $\varepsilon_O = -11.8$ eV for oxygen atoms.
For the nondiagonal matrix elements  $H^b_{ij}$ ($i \ne j$)
referring to neighboring $sp^2$-bonded
atoms we use the following expression proposed by Harrison \cite{harrison}:
\begin{equation}\label{harrison-pppi}
H^b_{ij} = V_{pp\pi} = -0.63 \frac{\hbar^2}{md^2},
\end{equation}
where $m$ is the electron mass and $d$ is the distance between the
corresponding nearest-neighboring atoms.
All other nondiagonal matrix elements,
referring to atoms not $sp^2$-bonded are assumed equal to zero, $H^b_{ij}=0$.

Diagonalizing the Hamiltonian matrix, one obtains $N$ molecular orbitals and their
eigenenergies. The lower energy orbitals are successively filled by
two electrons each, until to accommodate all available $p_z$ electrons.
Then the highest-energy occupied orbital, $\Psi^{b}_H ({\bf r})$, is the
$\pi$ HOMO and the lowest-energy unoccupied one, $\Psi^{b}_L ({\bf r})$,
is the $\pi$ LUMO. Following the usual convention,
we denote occupied molecular orbitals by $\pi$ and unoccupied ones by $\pi^*$.

\subsection{HOMO and LUMO of B-DNA base-pairs}
\label{sec:base-pairs}

Regarding the B-DNA base-pairs, we follow a different procedure in order
to obtain the HOMO and LUMO wavefunctions.
This is because the two bases (A and T, or G and C) are connected with
non-covalent hydrogen bonds to form the base-pair (A-T or G-C, respectively).
The length of hydrogen bonds, around 3 \AA, is longer than the typical
length of a covalent bond connecting neighboring atoms within a base,
which is around 1.3-1.5 \AA. In other words, a base-pair is not assumed
in our calculation like a single molecule but, instead, it
is treated like two adjacent molecules with electronic overlap.
However, we still use the terms HOMO and LUMO for the base-pairs,
meaning the single-electron wavefunctions which represent the highest in energy
occupied orbital and the lowest in energy unoccupied orbital, respectively,
of the molecular complex. We assume that these wavefunctions describe an
inserted hole or electron, respectively, within a base-pair.
Below we follow a straightforward linear combination of molecular orbitals
approach to designate the base-pairs' HOMO and LUMO.
Namely, the base-pair HOMO/LUMO ($H/L$) wavefunction reads
\begin{equation}\label{lcmo-base-pair}
\Psi^{bp}_{H/L}({\bf r}) = \mathcal{C}_1 \; \Psi_{H/L}^{b (1)} ({\bf r})
\; + \; \mathcal{C}_2 \; \Psi_{H/L}^{b (2)} ({\bf r}),
\end{equation}
where $\Psi_{H/L}^{b (1)} ({\bf r})$, $\Psi_{H/L}^{b (2)} ({\bf r})$
are the corresponding HOMO/LUMO orbitals of the two bases $(1)$ and $(2)$
forming the base-pair.
Inserting Eq. (\ref{lcmo-base-pair}) into Schr\"{o}dinger equation
$H^{bp} \; \Psi^{bp}_{H/L}({\bf r})= E^{bp}_{H/L} \; \Psi^{bp}_{H/L}({\bf r})$
(where $E^{bp}_{H/L}$ denotes the HOMO/LUMO base-pair energy),
multiplying once with $\Psi_{H/L}^{b (1) \; \star} ({\bf r})$ and once
with $\Psi_{H/L}^{b (2) \; \star} ({\bf r})$,
and integrating over all space, using
$\Psi_{H/L}^{b (1)} ({\bf r}) = \sum_{i=1}^{N_1} c_{i(1)}^{H/L}
p_z^{i (1)} ({\bf r})$,
$\Psi_{H/L}^{b (2)} ({\bf r}) = \sum_{j=1}^{N_2} c_{j(2)}^{H/L}
p_z^{j (2)} ({\bf r})$ [cf. Eq.~(\ref{lcao-pi-molecular})],
we obtain the following system of equations
\begin{eqnarray}
E_{H/L}^{b(1)} \; \mathcal{C}_1 & + \; t_{H/L} \; \mathcal{C}_2 & =
E^{bp}_{H/L} \; \mathcal{C}_1  \nonumber \\
t_{H/L}^\star \; \mathcal{C}_1 & + \; E_{H/L}^{b(2)} \; \mathcal{C}_2 & =
E^{bp}_{H/L} \; \mathcal{C}_2 \; .
\label{lcmosys}
\end{eqnarray}
Here we have assumed that $p_z$ orbitals belonging to atoms of
different bases are orthogonal and also that
$\int d{^3r} \; \Psi_{H/L}^{b (m) \; \star} \; H^{bp} \; \Psi_{H/L}^{b (m)}
\approx \int d{^3r} \; \Psi_{H/L}^{b (m) \; \star} \; H^b \;
\Psi_{H/L}^{b (m)}=E_{H/L}^{b (m)}$, for $m$=1 or 2.
The overlap integral,
$t_{H/L} = \int d{^3r} \; \Psi_{H/L}^{b (1) \; \star} \; H^{bp} \; \Psi_{H/L}^{b (2)}$,
which expresses
the hopping parameter for a charge (hole/electron) transfer between
the two bases of the base-pair, equals
\begin{equation}\label{tme}
t_{H/L} = \sum_{i=1}^{N_1} \sum_{j=1}^{N_2} c_{i(1)}^{H/L \; \star}
\; c_{j(2)}^{H/L} \; V_{ij},
\end{equation}
where $V_{ij} = \int d{^3r} \; {p_z^{i (1) \; \star}({\bf r})} \; H^{bp}
\; p_z^{j(2)} ({\bf r})$.

Because the base wavefunctions $\Psi^b ({\bf r})$ are real,
the coefficients $c_i$ in Eq.~(\ref{lcao-pi-molecular}) are real, and
the same holds for the overlap integrals $t_{H/L}$, i.e.,
$t_{H/L}^\star = t_{H/L}$. The matrix elements $V_{ij}$ are generally
provided through the Slater-Koster expression \cite{slater_koster_1954}:
\begin{equation} \label{s-k}
V_{ij}= V_{pp\sigma}  \; \mathrm{sin}^2 \phi + V_{pp\pi} \; \mathrm{cos}^2 \phi ,
\end{equation}
where $\phi$ denotes the angle formed by the line connecting atoms $i$ and $j$
and the plane perpendicular to $p_z$ orbitals (i.e. the plane of bases).
More details regarding the meaning of $V_{pp\sigma}$ and $V_{pp\pi}$
matrix elements can be found in Ref. \cite{harrison}.

For atoms belonging to different bases within a base-pair, i.e. the case
of interest in this subsection, the angle $\phi =$ 0 and $V_{ij}=V_{pp\pi}$.
Note that for intra-base covalently bonded neighboring atoms, $V_{pp\pi}$ are
the interatomic matrix elements proposed by Harrison \cite{harrison},
cf. Eq.~(\ref{harrison-pppi}). However, Harrison's relations are valid for
interatomic distances of the order of those of covalent bonds.
When dealing with larger interatomic distances, e.g. between atoms
belonging to different molecules, the proportional to
$1/d^2$ expressions of Harrison \cite{harrison} are replaced by appropriate
exponentially decaying expressions of the form \cite{MA,LA}
\begin{equation}  \label{lath}
V_{pp\pi}= A e^{-\beta (d-d_0)},
\end{equation}
where the constants $A$, $\beta$ are determined through the requirement
that at a typical covalent bond distance, $d_0$, the values of this expression and its
derivative in respect to $d$, coincide with those of Harrison's expressions
given in Eq. (\ref{harrison-pppi}). This yields $A=-0.63 \hbar^2 /md_0^2$
and $\beta=2/d_0$. Here we choose $d_0 = 1.35$ \AA.

Using Eq.(\ref{lath}) for $V_{ij}$ (since $V_{ij}$ is equal to $V_{pp\pi}$ for atoms
of different bases within the same base-pair, i.e. $\phi=0$)
and the known coefficients $c_i^{H/L}$ of the base HOMO/LUMO
(obtained as described in the previous subsection),
then $t_{H/L}$ are calculated through Eq.~(\ref{tme}). The quantities
$E_{H/L}^{b(1)}$ and $E_{H/L}^{b(2)}$ in Eq.~(\ref{lcmosys}) are
the HOMO/LUMO eigenenergies of the corresponding bases.
Therefore, the $2 \times 2$ system of Eq.~(\ref{lcmosys}) can be analytically
solved to determine $\mathcal{C}_1$, $\mathcal{C}_2$, and $E_{H/L}^{bp}$.
To obtain the HOMO (LUMO) of the base-pair, the higher (lower)
energy solution of the $2 \times 2$ system (\ref{lcmosys}) is considered
and $E_H^{bp}$ ($E_L^{bp}$) is the corresponding eigenenergy.
From the values $\mathcal{C}_1$ and $\mathcal{C}_2$ of the solution,
the base-pair HOMO/LUMO wavefunction is obtained through
Eq.~(\ref{lcmo-base-pair}).
For later use it is mentioned that by multiplying $\mathcal{C}_1$ and
$\mathcal{C}_2$ with the LCAO coefficients of the corresponding base
wavefunctions $\Psi_{H/L}^{b (1)}$ and $\Psi_{H/L}^{b (2)}$
[cf. Eq.~(\ref{lcao-pi-molecular})], the base-pair wavefunction
(\ref{lcmo-base-pair}) can be equivalently written as
\begin{equation}  \label{bpwf-pz}
\Psi^{bp}_{H/L}({\bf r}) = \sum_{i=1}^{N} C_{i}^{H/L} p_z^i ({\bf r}).
\end{equation}
Here the sum is extended over the $N$ atoms (contributing $p_z$ electrons
in $\pi$ bonds) of the whole base-pair ($N=18$ for A-T, while $N=19$ for G-C).

\subsection{Tight-binding parameters for charge transfer in B-DNA}
\label{sec:charge-transfer}

HOMO and LUMO energies of bases (or base-pairs) as well as
hopping parameters between successive bases (or base-pairs)
--calculated through a similar procedure like in the previous subsection,
cf. Eq.~(\ref{tme})--
provide an estimate of the parameters used in tight-binding models that
describe charge transport along DNA.
Such phenomenological models facilitate larger scale simulations,
reaching a mesoscopic level of description.
Depending on the particular problem of interest, a tight-binding description
at the base-pair level, or at the single-base level, may be more appropriate.
The relevant parameters for both cases are discussed below and
their estimated values are presented in Section III.

\subsubsection{Description at the base-pair level}
\label{sec:bpdes}

A tight-binding description of charge transfer between successive base-pairs
$\ldots, \lambda-1, \lambda, \lambda+1, \ldots$ of double-stranded DNA
can be obtained considering that extra electrons inserted in DNA travel
through LUMOs, while inserted holes travel through HOMOs.
In this approximation the time-dependent single carrier (hole/electron)
wavefunction of the whole macromolecule, $\Psi^{DNA}_{H/L}({\bf r},t)$,
is considered as a linear combination of base-pair wavefunctions with
time-dependent coefficients, i.e.,
\begin{equation}\label{psi-total}
\Psi^{DNA}_{H/L}({\bf r},t) = \sum_\lambda A_{\lambda}(t) \;
\Psi^{bp(\lambda)}_{H/L}({\bf r}),
\end{equation}
where $\Psi^{bp(\lambda)}_{H/L}({\bf r})$ is the $\lambda^{th}$
base-pair's HOMO/LUMO wavefunction and the sum is extended over all
base-pairs of the DNA molecule under consideration.

Starting from the time-dependent Schr\"{o}dinger equation,
$i \hbar \frac{d \Psi^{DNA}_{H/L}}{dt} =H^{DNA}\Psi^{DNA}_{H/L}$, and
following a similar procedure and assumptions similar to the ones of the
previous subsection,
one obtains that the time evolution of the coefficients $A_\lambda(t)$
obeys the tight-binding equations
\begin{equation}\label{TBbp}
i \hbar \frac{ dA_{\lambda}} {dt} = E^{bp (\lambda)}_{H/L} A_{\lambda}
+ t^{bp (\lambda;\lambda-1)}_{H/L} A_{\lambda-1}
+ t^{bp (\lambda;\lambda+1)}_{H/L} A_{\lambda+1}.
\end{equation}
Here, $E^{bp (\lambda)}_{H/L}$ is the HOMO/LUMO energy of base-pair
$\lambda$, as obtained in the previous subsection~\ref{sec:base-pairs}.
Depending on the actual sequence of the DNA molecule under consideration,
two values are possible for the on-site energies $E^{bp (\lambda)}_{H/L}$
of Eq.~(\ref{TBbp}), corresponding to A-T or G-C base-pairs.
Using the HOMO or LUMO base-pair wavefunctions ($\Psi_H^{bp}$ or
$\Psi_L^{bp}$), cf. Eq. (\ref{bpwf-pz}), the corresponding hopping parameters
($t_H^{bp}$ or $t_L^{bp}$) between successive base-pairs are obtained by
\begin{equation}\label{tibp}
t^{bp (\lambda;\lambda^\prime)}_{H/L} = \sum_{i=1}^{N_{\lambda}}
\sum_{j=1}^{N_{\lambda^\prime}} C_{i(\lambda)}^{H/L \; \star} \;
C_{j(\lambda^\prime)}^{H/L} \; V_{ij},
\end{equation}
where the indices $\lambda,\lambda^\prime$ denote neighboring base-pairs and
$V_{ij} = \int d{^3r} \; {p_z^{i(\lambda)\star}({\bf r})} \; H^{DNA} \;
p_z^{j(\lambda^\prime)}({\bf r})$. The sums over $i$ and $j$ in
Eq.~(\ref{tibp}) extend up to the total number of atoms $N_{\lambda}$ and
$N_{\lambda^\prime}$, respectively,
constituting the corresponding base-pair [cf. Eq.~(\ref{bpwf-pz})].
Here there is a difference with the hopping integrals in Eq.~(\ref{tme}), since in
the latter case the sums up to $N_1$ and $N_2$ extend over the total number
of atoms constituting the corresponding bases. The matrix elements $V_{ij}$ in
Eq.~(\ref{tibp}) are given by the semi-empirical Slater-Koster expression, Eq.~(\ref{s-k}).
Now $\phi \neq 0$ and  $V_{pp\pi}$ is obtained through Eq.~(\ref{lath}), as before.
$V_{pp\sigma}$ is also obtained from the same expression, Eq.~(\ref{lath}) but with a
different preexponential coefficient, $A=2.22 \hbar^2 /md_0^2$, resulting from the
constant 2.22 which appears in the corresponding to Eq.~(\ref{harrison-pppi})
Harrison's formula for $V_{pp\sigma}$~\cite{harrison}.
Therefore, using standard expressions from solid state physics
\cite{harrison,slater_koster_1954,MA,LA}, we are able to calculate the interatomic
matrix elements $V_{ij}$. The Slater-Koster expression of Eq. (\ref{s-k}) (see Table I of
Ref.~\cite{slater_koster_1954}) with matrix elements $V_{pp\pi}$ and $V_{pp\sigma}$
of the Harrison's type~\cite{harrison}, is typically used in solid state physics and it has
been also applied in the case of DNA charge transfer in Ref.~\cite{endres_et_al_2004}.
Our methodology for calculating the overlap matrix elements is very similar to that of the
latter work. The difference is that we use the exponentially decaying expressions
of the form of Eq. (\ref{lath}) for $V_{pp\pi}$ and $V_{pp\sigma}$, adopted from
Refs.~\cite{MA,LA}, instead of the fitting of similar expressions with {\it ab initio}
calculations as used in Ref.~\cite{endres_et_al_2004}.
All coefficients and matrix elements in Eq.~(\ref{tibp}) are real, resulting in real
$t^{bp(\lambda;\lambda^\prime)}_{H/L}=t^{bp(\lambda^\prime;\lambda)}_{H/L}$.

Knowledge of the coefficients $C_{i}^{H/L}$ from the base-pair wavefunctions
of Sec.~\ref{sec:base-pairs} [cf. Eq.~(\ref{bpwf-pz})] and of the matrix
elements $V_{ij}$ from the geometrical structure of DNA and the formulae
discussed above, allows the calculation of the hopping integrals
$t^{bp(\lambda;\lambda^\prime)}_{H/L}$ [cf. Eq.~(\ref{tibp})]
which appear as parameters in the tight-binding Eq.~(\ref{TBbp}).
Therefore, one can use the tight-binding parameters
$E^{bp (\lambda)}_{H/L}$ and $t^{bp(\lambda;\lambda^\prime)}_{H/L}$
computed in this work (cf. Table~\ref{table:bpHL} and
Table~\ref{table:interbpTP}, respectively, in Sec.~\ref{sec:resdis}),
in order to numerically solve the system of equations (\ref{TBbp}) and
obtain, through $A_{\lambda}(t)$, the time evolution of
a charge [cf. Eq. (\ref{psi-total})] propagating along any DNA segment.

\subsubsection{Description at the single-base level}

The tight-binding description at the single-base level is similar to
that in the previous subsection.
The difference is that the carrier wavefunction of the whole macromolecule
is now considered as a linear combination of single base wavefunctions,
instead of base-pair wavefunctions. Therefore
\begin{equation}\label{tb2wf}
\Psi^{DNA}_{H/L}({\bf r},t) = \sum_\lambda \left[ A_{\lambda}(t) \;
\Psi^{b(\lambda,1)}_{H/L}({\bf r}) \; + \; B_{\lambda}(t) \;
\Psi^{b(\lambda,2)}_{H/L}({\bf r}) \right],
\end{equation}
where $\lambda$ denotes base-pairs, the sum is again over all successive
base-pairs of DNA, and $\Psi^{b(\lambda,1)}_{H/L}$, $\Psi^{b(\lambda,2)}_{H/L}$
are HOMO/LUMO wavefunctions of bases in the $\lambda^{th}$ base-pair,
located at the one and the other DNA strands, respectively.

Under the same assumptions as previously and considering only
neighboring hoppings (between adjacent base-pairs), the tight-binding
equations for the time dependent coefficients of Eq. (\ref{tb2wf}) read
\begin{widetext}
\begin{eqnarray}
i \hbar \frac{dA_\lambda}{dt} = & E_{H/L}^{b(\lambda,1)} \; A_\lambda \;
+ \;  t_{H/L}^{b (\lambda,1;\lambda,2)} \; B_\lambda \; + \;
t_{H/L}^{b (\lambda,1;\lambda-1,1)} \; A_{\lambda-1} \; + \;
t_{H/L}^{b (\lambda,1;\lambda+1,1)} \; A_{\lambda+1}  \; + \;
t_{H/L}^{b (\lambda,1;\lambda-1,2)} \; B_{\lambda-1} \; + \;
t_{H/L}^{b (\lambda,1;\lambda+1,2)} \; B_{\lambda+1}     \nonumber \\
i \hbar \frac{dB_\lambda}{dt} = & E_{H/L}^{b(\lambda,2)} \; B_\lambda \;
+ \;  t_{H/L}^{b (\lambda,2;\lambda,1)} \; A_\lambda \; + \;
t_{H/L}^{b (\lambda,2;\lambda-1,2)} \; B_{\lambda-1} \; + \;
t_{H/L}^{b (\lambda,2;\lambda+1,2)} \; B_{\lambda+1} \; + \;
t_{H/L}^{b (\lambda,2;\lambda-1,1)} \; A_{\lambda-1} \; + \;
t_{H/L}^{b (\lambda,2;\lambda+1,1)} \; A_{\lambda+1} .
\label{TBb}
\end{eqnarray}
\end{widetext}
Here $E_{H/L}^b$ are base HOMO/LUMO energies and the hopping parameters
$t_{H/L}^b$ are interbase transfer integrals of the general form of
Eq.~(\ref{tme}) with $V_{ij}$ given by Eq.~(\ref{s-k}).
$\phi=0$ for interbase transfer integrals within the same base-pair
(interstrand intra-base-pair hoppings),
$t_{H/L}^{b (\lambda,1;\lambda,2)}=t_{H/L}^{b (\lambda,2;\lambda,1)}$,
while $\phi \neq 0$ for interbase transfer integrals between successive
bases within the same strand (intrastrand hoppings),
$t_{H/L}^{b (\lambda,i;\lambda \pm 1,i)}$, or between diagonally located, at
different strands, bases (interstrand inter-base-pair hoppings),
$t_{H/L}^{b (\lambda,i;\lambda \pm 1,j)}$ with $i \neq j$.
$V_{pp\pi}$ and $V_{pp\sigma}$ are given through exponentially decaying
Eq.~(\ref{lath}), as discussed before.

A tight-binding description at the single-base level requires solving of a
double number of differential equations, Eq.~(\ref{TBb}),
compared to the description at the base-pair level, Eq.~(\ref{TBbp}).
However, it allows individual base properties to be taken into account
when necessary. Section~\ref{sec:resdis} contains
the tight-binding parameters which appear in Eq.~(\ref{TBb});
see Table~\ref{table:bHL} for $E_{H/L}^b$,
Table~\ref{table:intrabpt} for $t_{H/L}^{b (\lambda,i;\lambda,j)}$,
Table~\ref{table:interbTP} for $t_{H/L}^{b (\lambda,i;\lambda\pm1,i)}$,
and Tables \ref{table:interbD3TP} and \ref{table:interbD5TP} for
$t_{H/L}^{b (\lambda,i;\lambda \pm 1,j)}$ with $i \neq j$.

\section{Results and Discussion}
\label{sec:resdis}

The results and their discussion have been organized in two subsections.
In subsection~\ref{sec:eigenenergies} we present our calculations for
the HOMO and LUMO energies as well as for the corresponding wavefunctions
of the four DNA bases, adenine, guanine, cytosine, and thymine, and
of the two base-pairs, adenine-thymine and guanine-cytosine.
We compare our results with available experimental data and
previous calculations using methods from first principles.

Subsection \ref{sec:hopping parameters} contains our calculations
for the charge transfer hopping parameters between successive base-pairs
(Sec.~\ref{sec:hopping parameters base-pairs}) or
between neighboring DNA bases (Sec.~\ref{sec:hopping parameters bases}),
and provides estimates for the corresponding tight-binding parameters
described previously in Section \ref{sec:charge-transfer}.
We present results for both electrons and holes and
for all possible combinations of successive bases or base-pairs.

\subsection{HOMO and LUMO energies and wavefunctions}
\label{sec:eigenenergies}

\begin{figure}
\epsfig{file=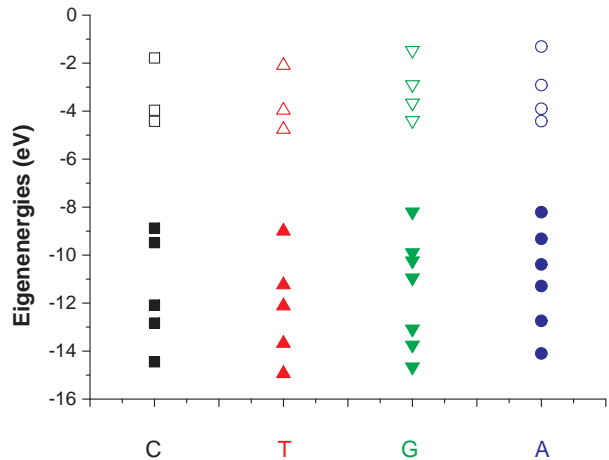,width=9cm}
\caption{(Color online) $\pi$ molecular structure of the four DNA bases:
cytosine (C), thymine (T),  guanine (G), and  adenine (A).
Full (empty) symbols
correspond to occupied (unoccupied) molecular orbitals.}
\label{bener}
\end{figure}

Before starting our discussion about the HOMO and LUMO of DNA bases,
we show in Fig.~\ref{bener} the $\pi$ electronic structure of all bases,
as obtained using the LCAO method presented in Sec.~\ref{sec:bases}.
The geometrical structure of bases and the corresponding interatomic
distances have been obtained from Ref. \cite{gstruct}.
Occupied orbitals are shown by full symbols in Fig.~\ref{bener},
while unoccupied ones by empty symbols.
The second and third row of Table~\ref{table:bHL} list --for each base of
the corresponding column-- the number of atoms participating in $\pi$-bonds
and the total number of $p_z$ electrons in the molecule, respectively.
For example, adenine has twelve $p_z$ electrons, which occupy, by pairs
of opposite spin, the six lowest energy levels of A shown in Fig.~\ref{bener}.

\begin{table}
\caption{For each DNA base (first row) are shown:
the number of atoms participating in $\pi $ bonds (second row),
the total number of contributed $p_z$ electrons (third row),
the $\pi$HOMO energy $E_H^{b \; (isol.)}$ (fourth row) and
the $\pi$LUMO energy $E_L^{b \; (isol.)}$ (fifth row) of the isolated
base, as well as the energy of the first $\pi$-$\pi^*$ transition
$E_{\pi-\pi^*} = E_L^{b \; (isol.)} - E_H^{b \; (isol.)}$ (sixth row),
obtained in this work.
We also show experimental values of the ionization energy $IE$
(seventh row) and the first $\pi$-$\pi^*$ transition (eighth row),
as well as existing theoretical predictions using methods from first
principles for $IE$ (ninth row) and $E_{\pi-\pi^*}$ (tenth row).
Finally, we list the $\pi$HOMO and $\pi$LUMO energies, $E_H^b$ (eleventh row)
and $E_L^b$ (twelfth row), respectively, of the bases when they are distorted
within the base-pairs of B-DNA (see text), as obtained in our work.
The quantities $E_H^b$ in eleventh row and $E_L^b$ in twelfth row represent
the parameters $E_{H/L}^{b(\lambda,i)}$ which appear in Eq.~(\ref{TBb}).
All energies are given in eV.}
\centerline{
\begin{tabular}{|l|c|c|c|c|} \hline
DNA base& Adenine & Thymine & Guanine & Cytosine \\  \hline  \hline
atoms in $\pi$ bonds&   10  &     8 &    11 &     8  \\ \hline
$p_z$ electrons     &   12  &    10 &    14 &    10  \\ \hline
$E_H^{b \; (isol.)}$& $-$8.2& $-$9.0& $-$8.2& $-$8.9 \\ \hline
$E_L^{b \; (isol.)}$& $-$4.4& $-$4.8& $-$4.4& $-$4.4 \\ \hline
$E_{\pi-\pi^*}$     &   3.8 &    4.2&    3.8&    4.5 \\ \hline
$IE^{\mathrm{exp}}$ &       &       &       &        \\ \cite{HC Chem. Phys. Let 1975, Tetrahedron 45_1975,DWMG, JACS 10_1980 and JPC 84_1980, JMolSt 214_1989}
& 8.4-8.5 & 9.0-9.2 & 8.2-8.3 & 8.9 \\ \hline
$E_{\pi-\pi^*}^{\mathrm{exp}}$ & &  &       &        \\ \cite{VGCD_Biopolymers_1963,YF_Biopolymers_1968, MN,Clark_JPC_1990,VRBD_JACS_1968,MRE_PNAS_67,CPT_JPC_1965,
SJ_Biopolymers_1977,HBAN_JACS_97,BM_BIOPOLYMERS_75,RFFK_Biopolymers_1978, ZNC_JACS_1985,FF_JACS_1971} 
& 4.5-4.8 & 4.6-4.7 & 4.3-4.5 & 4.5-4.7  \\ \hline
$IE^{\mathrm{first \; pr.}}$ & &    &       &\\ \cite{SS_JACS_1996,CBCS_JPC_1992,HC_JACS_1996}
 & 8.2-8.6 & 9.1-9.7 & 7.8-8.3 & 8.9-9.4 \\ \hline
$E_{\pi-\pi^*}^{\mathrm{first \; pr.}}$ & & & & \\ 
\cite{FSR_JACS_1997,MTT_JPCA_2001,SD_EPJD_2002,SL_JCC_2004,LFR_JACS_1995,SM_CP_1999,IBOKR_JACS_2002,SL_JPHA_2002,FR_JACS_1995,VFMR_JPCB_2006,LCS_JPCA_and_JPCB_01_02,FKH_JPCA_2007}
& 4.5-5.3 & 4.9-5.3 & 4.4-5.3 & 4.2-4.7 \\ \hline
$E_H^b$ (in B-DNA) &$-$8.3& $-$9.0& $-$8.0& $-$8.8 \\ \hline
$E_L^b$ (in B-DNA) &$-$4.4& $-$4.9& $-$4.5& $-$4.3 \\ \hline
\end{tabular}  }\label{table:bHL}
\end{table}

Table~\ref{table:bHL} summarizes our results for the HOMO and LUMO
energies of DNA bases. Relevant experimental values for the vertical
ionization energy (which is equal to the absolute value of HOMO energy) and
the first $\pi-\pi^*$ transition (i.e. the energy difference between
LUMO and HOMO) are also shown, along with corresponding results
of previous theoretical calculations using methods from first principles.
When the bases form base-pairs within DNA, they are slightly deformed
in comparison to their structure when isolated.
The HOMO and LUMO energies of the distorted bases in B-DNA
(shown in the last two rows of Table~\ref{table:bHL})
may differ from those of the isolated bases, and these values provide
the relevant on-site energies $E_{H/L}^b$ in the tight-binding Eq. (\ref{TBb}).

\begin{table}
\caption{For the two B-DNA base-pairs (first row) are shown:
the $\pi$HOMO energy $E_H^{bp}$ (second row), the $\pi$LUMO energy $E_L^{bp}$ (third row),
and the corresponding first $\pi$-$\pi^*$ transition energy
$E_{\pi-\pi^*}=E_L^{bp}-E_H^{bp}$ (fourth row), as obtained in this work.
We also list existing theoretical predictions using methods from first principles,
for the HOMO energy (fifth row) and the first $\pi$-$\pi^*$ transition (sixth row).
The quantities $E_H^{bp}$ in the second row and $E_L^{bp}$ in the third row
represent the parameters $E_{H/L}^{bp(\lambda)}$ which appear in Eq.~(\ref{TBbp}).
All energies are given in eV.}
\centerline{
\begin{tabular}{|l|c|c|} \hline
B-DNA base-pair     & A-T     &   G-C    \\  \hline  \hline
$E_H^{bp}$       & $-$8.3 & $-$8.0  \\  \hline
$E_L^{bp}$       & $-$4.9 & $-$4.5  \\  \hline
$E_{\pi-\pi^*}$  & 3.4 &   3.5  \\  \hline
$E_H^{bp \; \mathrm{first \; pr.}}$
\cite{SS_JACS_1996,HC_JACS_1996,ZLYY_JCP_2002,LCS_JPCA_and_JPCB_01_02,SL_JPCA_2002}
& $-$(7.8-8.2) & $-$(6.3-7.7)  \\ \hline
$E_{\pi-\pi^*}^{\mathrm{first \; pr.}}$ \cite{SL_JPCA_2002,VFMR_JPCB_2006}
& 6.4  & 4.3-6.3   \\ \hline
\end{tabular}  } \label{table:bpHL}
\end{table}

In Table~\ref{table:bpHL} we report results for the HOMO and LUMO energies
(and the HOMO-LUMO energy gap) of the two B-DNA base-pairs, using the
procedure desribed in Section~\ref{sec:base-pairs}. Corresponding values
obtained previously from first principles methods are also shown.

\begin{figure}
\epsfig{file=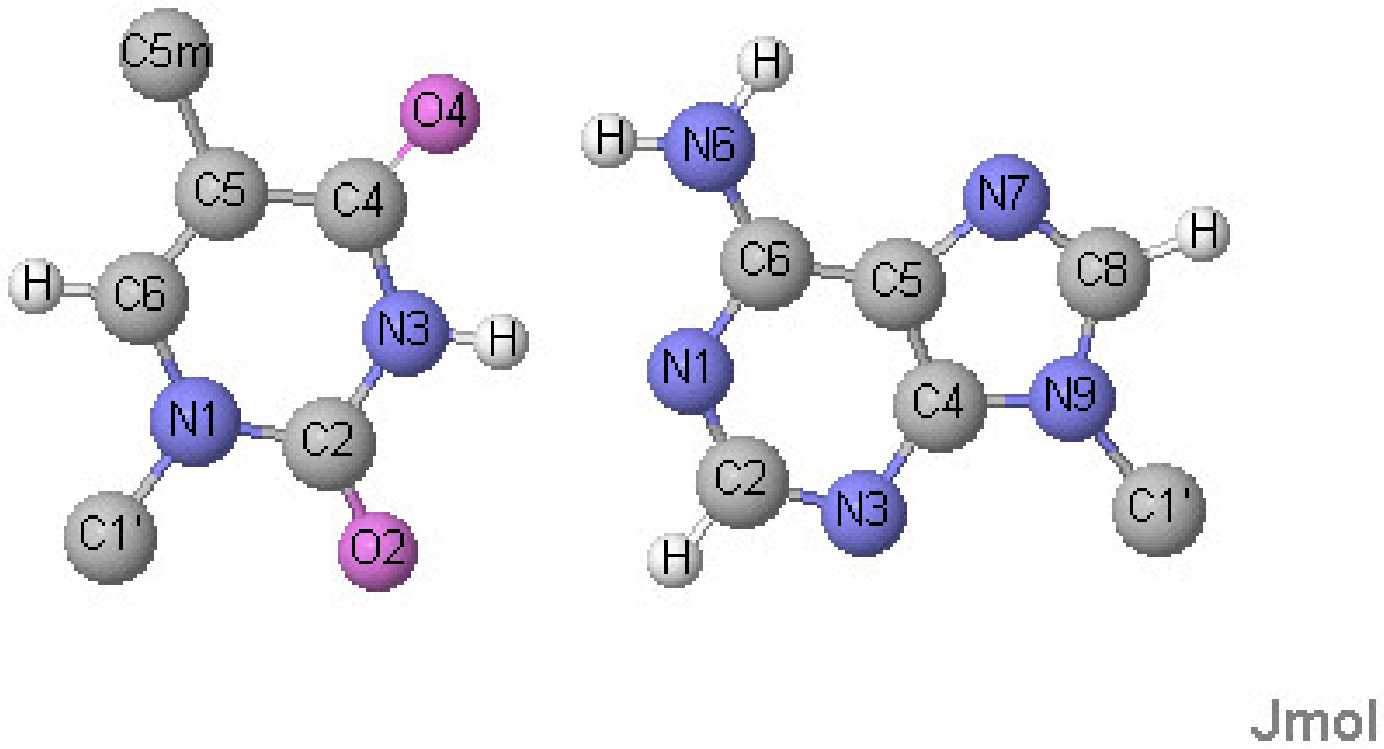,width=0.87\linewidth}
\epsfig{file=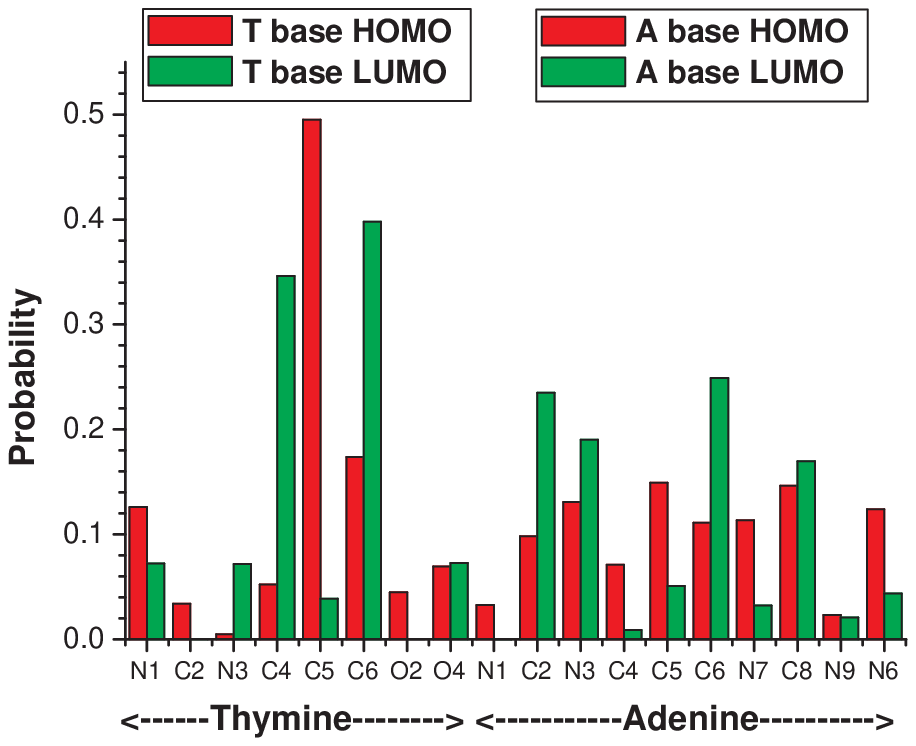,width=0.9\linewidth}
\epsfig{file=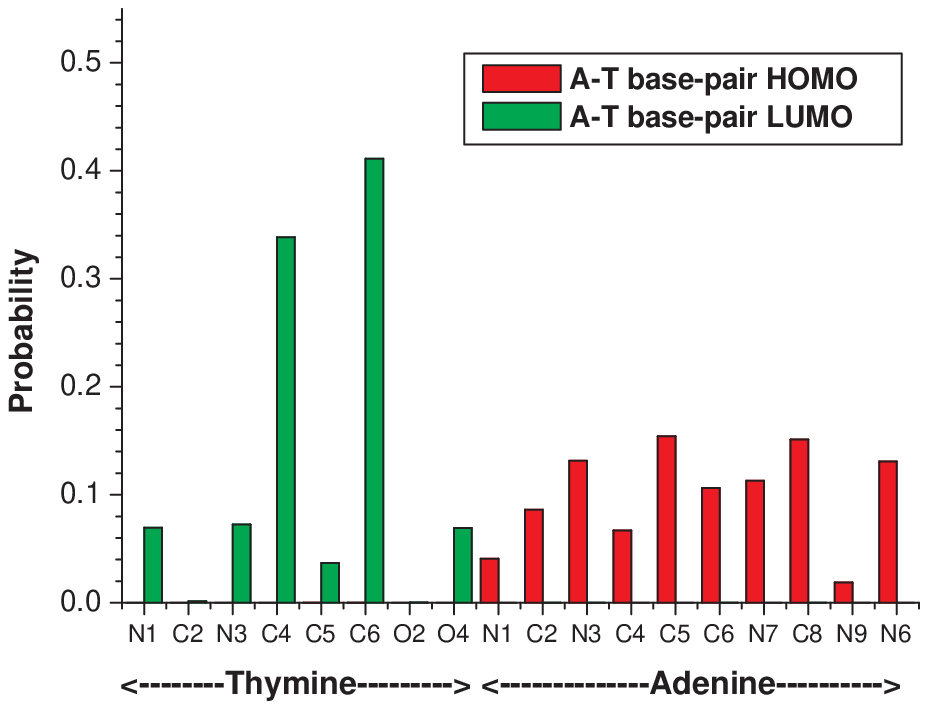,width=0.9\linewidth}
\caption{ (Color online) {\it Top:} geometrical structure of adenine (right) and
thymine (left), within an A-T base-pair. It is shown
the standard numbering of atoms, as also used in this work.
C1$'$ denote deoxyribose carbons and C5m thymine's methyl carbon, not
participating in $\pi$ bonding.
{\it Middle:} atomic occupation probabilities, $|c_i|^2$ [cf. Eq.~(\ref{lcao-pi-molecular})],
for the HOMO and LUMO wavefunctions of isolated adenine (right) and
thymine (left) bases.
{\it Bottom:} atomic occupation probabilities, $|C_i|^2$ [cf. Eq.~(\ref{bpwf-pz})],
for the HOMO and LUMO wavefunctions of A-T base-pair.}
\label{figAT}
\end{figure}

\begin{figure}
\epsfig{file=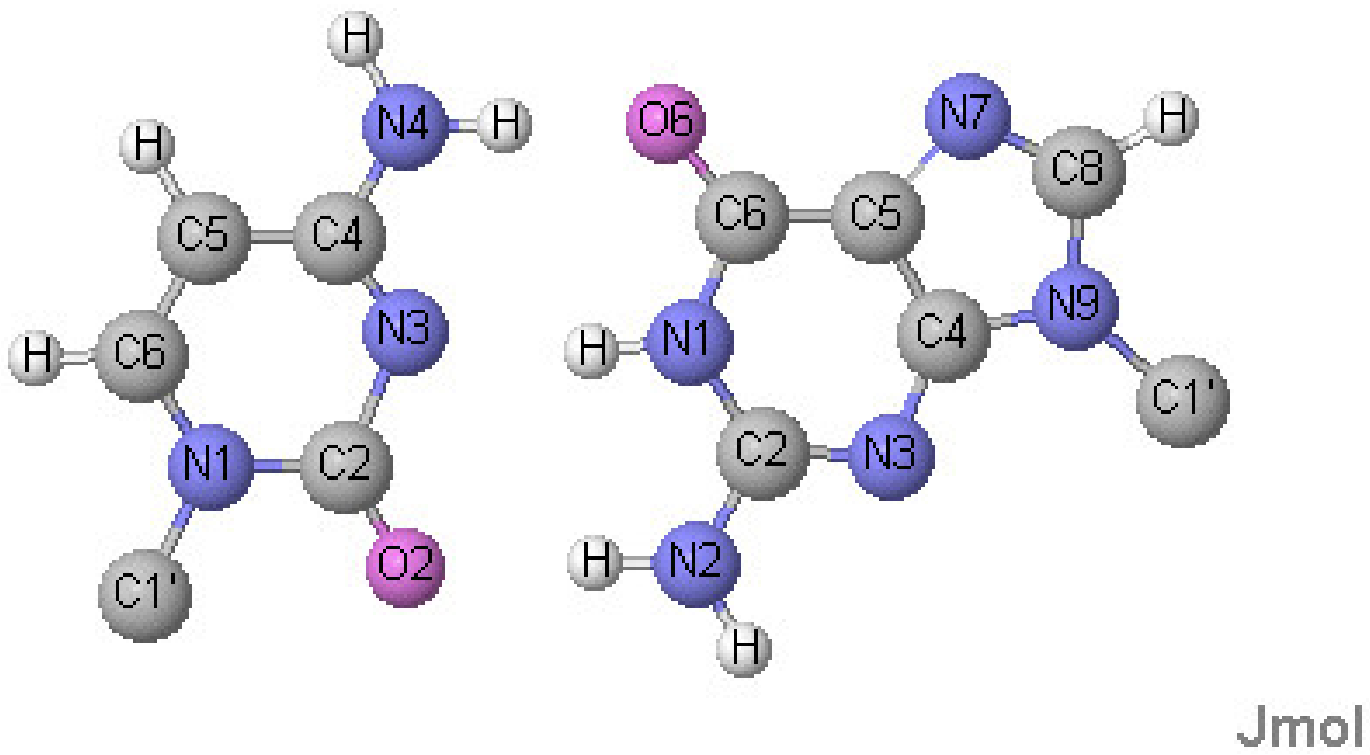,width=0.87\linewidth}
\epsfig{file=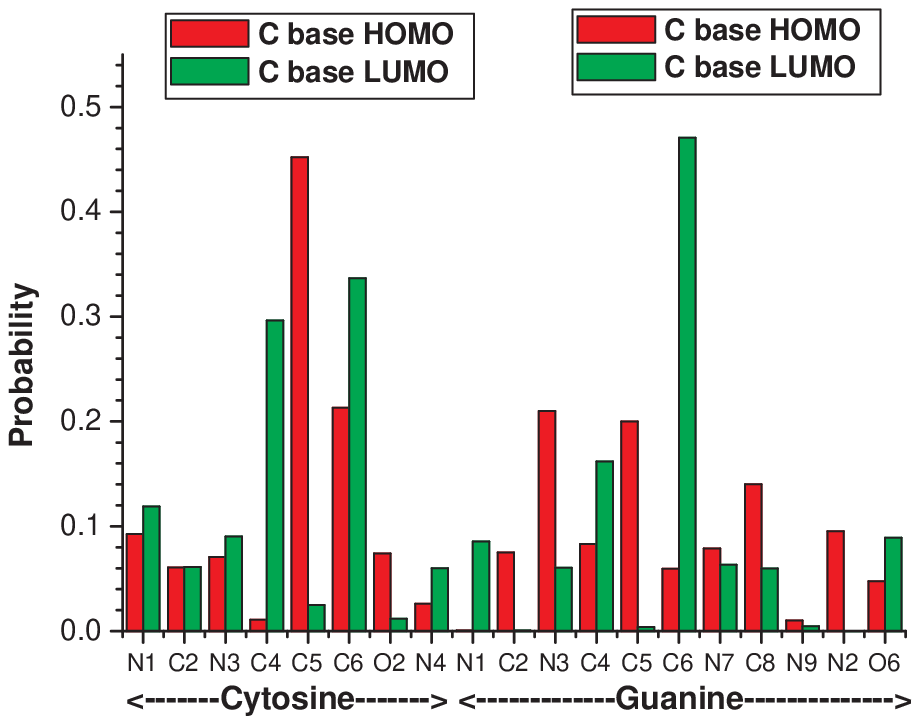,width=0.9\linewidth}
\epsfig{file=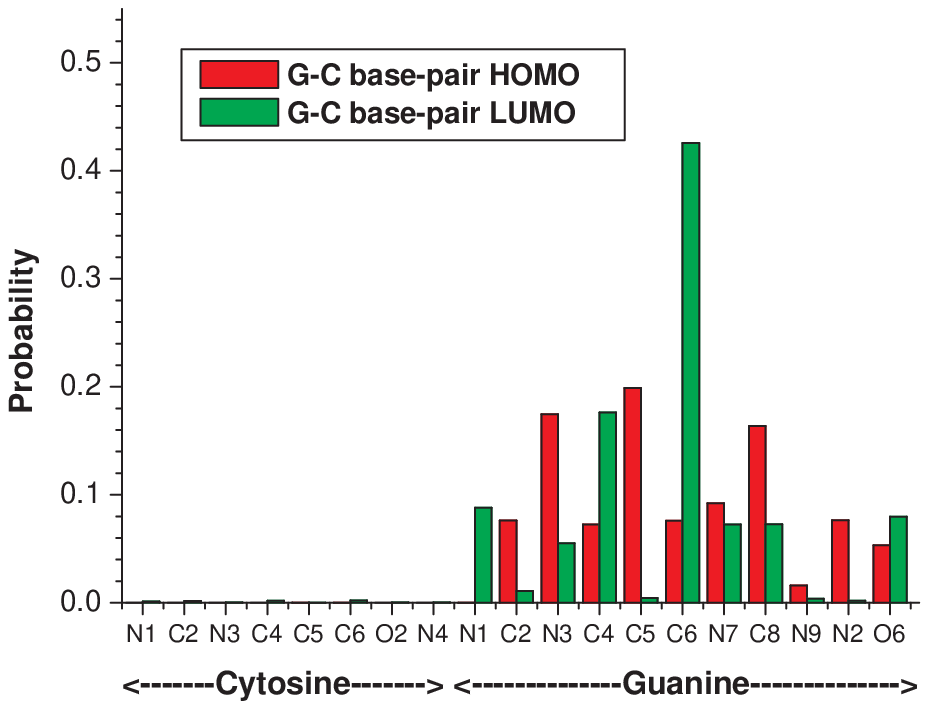,width=0.9\linewidth}
\caption{ (Color online) {\it Top:} geometrical structure of guanine (right) and
cytosine (left), within a G-C base-pair. It is shown
the standard numbering of atoms, as also used in this work.
C1$'$ denote deoxyribose carbons, not participating in $\pi$ bonding.
{\it Middle:} atomic occupation probabilities, $|c_i|^2$ [cf. Eq.~(\ref{lcao-pi-molecular})],
for the HOMO and LUMO wavefunctions of isolated guanine (right) and
cytosine (left) bases.
{\it Bottom:} atomic occupation probabilities, $|C_i|^2$ [cf. Eq.~(\ref{bpwf-pz})],
for the HOMO and LUMO wavefunctions of G-C base-pair.}
\label{figGC}
\end{figure}

Figure~\ref{figAT} (Figure~\ref{figGC}) presents for A, T, and A-T
(for G, C, and G-C) the corresponding intramolecular occupation probabilities
of holes or electrons, obtained from the calculated HOMO or LUMO
wavefunctions, respectively. This is shown through
the squared coefficients $|c_i|^2$ (or $|C_i|^2$ for base-pairs)
in the expansion of Eq.~(\ref{lcao-pi-molecular})
(or Eq.~(\ref{bpwf-pz}) for base-pairs) of the corresponding wavefunction.
These squared coefficients represent the probabilities of finding an
electron or hole (LUMO or HOMO, respectively)
located at the corresponding atom of the base or the base-pair.
The structures of the bases, prepared with the open-source software
for chemical structures Jmol \cite{Jmol}, are shown on the top of each
figure. The standard numbering of atoms has been used.

A detailed discussion regarding the HOMO and LUMO wavefunctions
is carried out below, separately for each isolated base and for the
two B-DNA base-pairs.

\subsubsection{Adenine}
\label{subsubsec:adenine}

Our calculations show that the HOMO energy of adenine is $-8.2$ eV.
This slightly underestimates the experimental value of the ionization
energy (8.4 eV \cite{HC Chem. Phys. Let 1975}, 8.5 eV
\cite{JACS 10_1980 and JPC 84_1980,JMolSt 214_1989}) by 0.2-0.3 eV.
Our result is similar to that of Ref.~\cite{SS_JACS_1996},
where using {\it ab initio} calculations the HOMO energy was found $-8.24$ eV.
Other {\it ab initio} calculations predicted the ionization energy between
8.3-8.6 eV \cite{CBCS_JPC_1992}. In good accordance with the measured values
is also the {\it ab initio} prediction of 8.5 eV in Ref.~\cite{HC_JACS_1996}.

The calculated HOMO wavefunction (see Fig.~\ref{figAT}, middle right)
is distributed all over adenine.
In descending order, it has its main amplitude in atoms
C5, C8, N3, N6, N7, C6, C2, C4,
while the lowest probabilities exist in atoms N1 and N9.
These results agree with those from Ref.~\cite{fuentes-cabrera_et_al},
since in both cases the HOMO wavefunction is distributed all over
adenine and the lowest probabilities are found at the same atoms.
Our results are also in agreement with the outcome of density functional
theory calculations \cite{CKK_JPCB}, where the $\pi$ HOMO wavefunction
was found to be delocalized along adenine, with greater probabilities
in atoms N7, N3, C6, C5, C8, and the lowest in atom N9, as well as
with recent time-dependent density functional theory predictions
\cite{VFMR_JPCB_2006}, where the HOMO wavefunction is distributed
all over the molecule except from atom N9.

We find the LUMO energy of adenine at $-4.4$ eV.
According to our calculations, the first $\pi$-$\pi^*$ transition energy
is $E_{\pi-\pi^*} = 3.8$ eV. The deviation from the experimental value (4.5-4.8 eV
\cite{VGCD_Biopolymers_1963,YF_Biopolymers_1968,Clark_JPC_1990,VRBD_JACS_1968,
SJ_Biopolymers_1977,HBAN_JACS_97,BM_BIOPOLYMERS_75}) is 0.7-1.0 eV.
The $E_{\pi-\pi^*}$ value reported from several first principles methods
is around 4.5-5.3 eV
\cite{FSR_JACS_1997,MTT_JPCA_2001,SD_EPJD_2002,SL_JCC_2004,VFMR_JPCB_2006,FKH_JPCA_2007},
i.e. the corresponding deviation from the experimental observations is
0-0.8 eV.
It must be mentioned here that, although it is generally accepted
for all DNA bases that the HOMO-LUMO energy gap corresponds to the
first $\pi$-$\pi^*$ transition
\cite{HC Chem. Phys. Let 1975,JACS 10_1980 and JPC 84_1980,JMolSt 214_1989,VRBD_JACS_1968,SJ_Biopolymers_1977},
in Ref.~\cite{FKH_JPCA_2007} the HOMO-LUMO gap of adenine is attributed to
a $n$-$\pi^*$ transition, while in Refs. \cite{SL_JCC_2004,VFMR_JPCB_2006}
the $\pi$-$\pi^*$ and $n$-$\pi^*$ transitions are predicted to be very close.

Regarding the LUMO wavefunction of adenine, it can be seen from
Fig.~\ref{figAT} (middle right) that the highest probabilities are found
in atoms C6, C2, N3, and C8, while the remaining atoms are occupied
by much smaller probabilities ($\lesssim 5 \%$).
Our findings agree with those of Ref.~\cite{fuentes-cabrera_et_al},
where the LUMO wavefunction is localized in atoms C8, C6, C2, N3, and N7.
The only difference from our findings is the existence
of a relatively higher probability in atom N7. In both results the lowest
probabilities appear in the same atoms. Our findings are also in
good accordance with the results of Ref.~\cite{VFMR_JPCB_2006}, where
the LUMO wavefunction is localized in the same atoms (C8, C6, C2, N3).

\subsubsection{Thymine}
\label{subsubsec:thymine}

Thymine's HOMO energy is found $-9.0$ eV, which is in agreement with
the experimental value of ionization energy, $IE_\textrm{T}=$9.0-9.2 eV
\cite{HC Chem. Phys. Let 1975,DWMG,JMolSt 214_1989,Tetrahedron 45_1975}.
The {\it ab initio} predictions of Refs. \cite{SS_JACS_1996} and
\cite{HC_JACS_1996} fall also within the range of experimental results, as
they found the HOMO energy at $-9.14$ eV and $-9.16$ eV, respectively.
However, other {\it ab initio} calculations \cite{CBCS_JPC_1992}
give the ionization energy between 9.4-9.7 eV, which is in worse
agreement with the experimental data (a deviation of 0.2-0.7 eV).

According to our calculations the HOMO wavefunction of thymine
(cf. Fig.~\ref{figAT}, middle left) is mainly localized at the atom C5
(with almost 50$\%$ probability) and
at its nearby atoms C6 and N1 (with total probability around $30 \%$).
The remaining atoms have probabilities less than $10 \%$ each, with a
negligible probability at the atom N3.

We find that the LUMO energy of thymine is $-4.8$ eV.
Looking at the first $\pi$-$\pi^*$ transition energy, our prediction
of 4.2 eV underestimates the experimental results (4.6-4.7 eV
\cite{VGCD_Biopolymers_1963, YF_Biopolymers_1968,VRBD_JACS_1968,MRE_PNAS_67,
SJ_Biopolymers_1977,BM_BIOPOLYMERS_75,CPT_JPC_1965}) by 0.4-0.5 eV.
Calculated $E_{\pi-\pi^*}$ values from several {\it ab initio} methods are
between 4.9-5.3 eV \cite{SL_JCC_2004,LFR_JACS_1995,SM_CP_1999,FKH_JPCA_2007},
which overestimate the observed values by 0.2-0.7 eV.
We mention that Refs. \cite{SL_JCC_2004,LFR_JACS_1995,SM_CP_1999,FKH_JPCA_2007}
attribute the HOMO-LUMO gap of thymine to a $n$-$\pi^*$ transition.

Fig.~\ref{figAT} (middle left) shows that the LUMO wavefunction of thymine
exhibits its main amplitudes in atoms C6 and C4, with a total probability
around $75 \%$. All other atoms have less than $10 \%$ probability each,
with vanishing amplitudes at the atoms C2 and O2.
Our predictions agree with those from Ref.~\cite{VFMR_JPCB_2006},
where the LUMO wavefunction is mainly localized in atoms C4, C5, and C6.

\subsubsection{Guanine}
\label{subsubsec:guanine}

The HOMO energy of guanine is $-8.2$ eV according to our results. Our
prediction for its ionization energy almost coincides with the experimental
observations (8.24 eV \cite{HC Chem. Phys. Let 1975} and 8.28 eV
\cite {JACS 10_1980 and JPC 84_1980}).
Previous {\it ab initio} calculations have found the
ionization energy of guanine at 8-8.3 eV \cite{CBCS_JPC_1992},
7.75 eV \cite{SS_JACS_1996}, and 8.21 eV \cite{HC_JACS_1996}.
These values deviate from the experimental ones by 0-0.5 eV.

We observe in Fig.~\ref{figGC} (middle right) that the HOMO wavefunction
of guanine is extended over a large part of the molecule, containing the
atoms N2-C2-N3-C4-C5-N7-C8 (with total probability around $90 \% $).
The atoms N3 and C5 present the highest probabilities (around $20 \%$ each),
while about $15 \%$ probability exists in atom C8.
A vanishing probability appears at the atom N1 and a very small one at N9.
Our results agree with those of Ref.~\cite{CKK_JPCB}, where the main
amplitude appears in the part N2-C2-N3-C4-C5 of the molecule
and a lower probability also exists in atom C8.
The main difference between the two calculations is the existence of a probability
$18\%$ at the atom O6 in Ref.~\cite{CKK_JPCB}, compared to $5\%$ in our case.
Similar to our findings are the predictions obtained from the {\it ab initio}
method of Ref.~\cite{FCM_PRB}, where the HOMO wavefunction is mainly
localized at the same part of guanine, while there is also agreement with
the results of Ref.~\cite{VFMR_JPCB_2006}, where the HOMO is
predicted to be distributed all over the molecule apart from the atoms
N1, C6, and N9.

We find that Guanine's LUMO energy is $-4.4$ eV.
Regarding the first $\pi$-$\pi^*$ transition, our prediction is 3.8 eV,
which underestimates the observed values (between 4.3-4.5 eV
\cite{VGCD_Biopolymers_1963,YF_Biopolymers_1968,MN,VRBD_JACS_1968})
by 0.5-0.7 eV.
The first $\pi$-$\pi^*$ excitation energy obtained from several
first principles methods is between 4.4-5.3 eV
\cite{FSR_JACS_1997,SL_JCC_2004,VFMR_JPCB_2006,FKH_JPCA_2007},
presenting deviations of 0-0.8 eV from the experimental observations.

Looking at the LUMO wavefunction of guanine (Fig.~\ref{figGC}, middle right),
we see that the main amplitude appears in atom C6 with $\approx 45\%$
occupation probability. A significant probability also exists in atom C4 (more than
$15 \%$). All other atoms have probabilities less than $10 \%$,
with negligible amplitudes at C2, N2, C5, and N9. These results differ
from those of Ref.~\cite{FCM_PRB}, since there the LUMO wavefunction
exhibits higher probabilities at the atoms C2, C4, N1, C5, and N9, in
descending order. On the contrary, the present findings are in reasonable
agreement with those of Ref.~\cite{VFMR_JPCB_2006}.

\subsubsection{Cytosine}
\label{subsubsec:cytosine}

The HOMO energy of cytosine is found $-8.9$ eV. Our prediction for the
ionization energy coincides with the measured value of 8.94 eV
\cite{HC Chem. Phys. Let 1975}. Corresponding values calculated from
{\it ab initio} methods are 8.87 eV \cite{SS_JACS_1996}, 8.88 eV
\cite{HC_JACS_1996}, and 9.0-9.4 eV \cite{CBCS_JPC_1992}. The former
predictions estimate the experimental ionization energy equally well with
our simple LCAO calculations, while the latter ones worse.

It can be seen from Fig.~\ref{figGC} (middle left) that the HOMO
wavefunction of cytosine is mainly localized at C5 (occupation probability
$\approx 45 \%$), presenting also a significant amplitude at the
neighboring atom C6 (probability larger than $20 \%$).
Atoms N1, C2, O2, and N3 exhibit probabilities between $5\%-10\%$,
while C4 and N4 have the smaller probabilities (below $2 \%$ each).
In Ref.~\cite{CKK_JPCB} it was found that the atoms C5 and C6 accumulate
only $25\%$ probability, while the higher amplitudes appear in atoms
O2 and N3.

We find the LUMO energy of cytosine equal to $-4.4$ eV.
The first $\pi$-$\pi^*$ transition energy according to our results (4.5 eV)
is compared well with the experimentally observed values (4.5-4.7 eV
\cite{VGCD_Biopolymers_1963,YF_Biopolymers_1968,MN,VRBD_JACS_1968,SJ_Biopolymers_1977,RFFK_Biopolymers_1978,ZNC_JACS_1985,FF_JACS_1971,FKH_JPCA_2007}).
Results from first principles methods predict values between 4.2-4.7 eV,
i.e. some of them are within the range of experimental values
\cite{SL_JCC_2004,IBOKR_JACS_2002,SL_JPHA_2002}, while others are
slightly below the observed values \cite{FR_JACS_1995,VFMR_JPCB_2006}.

Figure \ref{figGC} (middle left) shows that the LUMO wavefunction of cytosine
is mainly localized in atoms C6 and C4 with occupation probabilities around
$30 \%$ each. A probability more than 10$\%$ appears in N1, while C2, N3,
 and N4 have probabilities between $5\%-10\%$. These findings are in
accordance with the results of Ref. \cite{VFMR_JPCB_2006}, where
the higher probabilities are found in atoms C4, C5, and C6.

\subsubsection{A-T base-pair}
\label{subsubsec:AT_base-pair}

As it has been already mentioned, the bases are slightly deformed within the
base-pairs of B-DNA. This distorted form of the bases is appropriate
when discussing base-pair properties (HOMO and LUMO energies or wavefunctions),
as opposed to the isolated form of the bases considered in the previous
subsections. Therefore, before going to the HOMO and LUMO discussion
of base-pairs, we briefly mention the changes in bases' HOMO and LUMO
due to the structural distortion within base-pairs.

For adenine in the B-DNA conformation the HOMO energy is $E_H^A=-8.3$ eV
(differing by $0.1$ eV in respect with the value of the isolated base, $-8.2$ eV,
see Table~\ref{table:bHL}), while the LUMO energy remains unchanged ($E_L^A=-4.4$ eV).
Regarding the HOMO and LUMO wavefunctions of adenine the results are
almost similar in both conformations. For example in the case of the HOMO
wavefunction, the highest difference of the occupation probabilities in respect with
the results of the isolated base (Fig.~\ref{figAT}, middle) is $ -1.2 \%$
in atom C2, while in all other atoms the differences are below $1\%$.
In the case of LUMO wavefunction the differences are slightly
higher, reaching $+4.1 \%$ in atom C8, $-3.3 \%$ in atom C2, $-2.4 \%$ in atom N3,
and $+2.1 \%$ in atom N7, while in the other atoms the changes are below $1.2 \%$.

For thymine the HOMO energy using the B-DNA structure is invariable
in respect with that of the isolated base ($E_H^T=-9.0$ eV), while the LUMO energy
changes slightly from $-4.8$ eV in the isolated form to $E_L^T-4.9$ eV
in the B-DNA conformation. The occupation probabilities for the HOMO (LUMO)
wavefunction of thymine are almost similar in both cases and the differences
are not larger than $0.5 \%$ ($1.5 \%$).

According to our method, the HOMO energy of A-T base-pair is
$E_H^{AT} = -8.3$ eV. Comparing with {\it ab initio} results, this value
differs only by $0.1$ eV with the prediction of Ref.~\cite{SS_JACS_1996}
($-8.19$ eV), it is very close to the ones calculated in Refs.
\cite{ZLYY_JCP_2002} ($-8.12$ eV) and \cite{HC_JACS_1996} ($-8.06$ eV),
while there is a reasonable agreement with the value given in
Ref.~\cite{LCS_JPCA_and_JPCB_01_02} ($-7.8$ eV).

The LUMO energy of A-T base-pair is $E_L^{AT} = -4.9$ eV. Therefore,
the first $\pi$-$\pi^*$ excitation energy in our calculations is 3.4 eV.
The prediction of $E_{\pi-\pi^*}$ obtained from first principles methods
in Ref. \cite{SL_JPCA_2002} is 6.39 eV.
It must be mentioned that according to the latter study
the HOMO-LUMO gap is not assigned to a $\pi$-$\pi^*$ transition,
but to a $\pi$-$\sigma^*$ transition~\cite{SL_JPCA_2002}.

Figure \ref{figAT} (bottom) shows that for the A-T base-pair the HOMO
wavefunction is completely localized in adenine, while, on the contrary,
the LUMO is completely localized in thymine.
This is justified because of the significantly higher HOMO energy
of A with respect to that of T ($E_H^A-E_H^T=0.7$ eV) and the
significantly lower LUMO energy of T with respect to that of A
($E_L^T-E_L^A=-0.5$ eV), respectively, as compared with the corresponding
overlap integrals of the bases' wavefunctions within the base-pair
($|t_H| = 12$ meV and $|t_L| = 9$ meV, see Table~\ref{table:intrabpt} below).
This also explains that $E_H^{AT}=E_H^A$ and $E_L^{AT}=E_L^T$, as well as that
the electron (hole) distribution within the base-pair is identical with the
corresponding distribution in thymine (adenine). The latter can be seen by a comparison
of the corresponding wavefunctions in the middle and bottom panels of Fig.~\ref{figAT},
taking also into account that adenine's HOMO and thymine's LUMO wavefunctions remain
practically the same in the B-DNA and isolated conformations, as it is mentioned above.

\subsubsection{G-C base-pair}
\label{subsubsec:GC_base-pair}

The HOMO energy of guanine base in the base-pair conformation is $E_H^G=-8.0$ eV.
There is a difference of $0.2$ eV compared to the corresponding result of the isolated
base ($-8.2$ eV). Its LUMO energy in the B-DNA conformation ($E_L^G=-4.5$ eV) changes
by $0.1$ eV in comparison to the value of $-4.4$ eV obtained for the isolated base.
Concerning the HOMO wavefunction of guanine the highest differences of the occupation
probabilities in respect with the results for the isolated base (shown in Fig.~\ref{figGC},
middle) are $-3.5 \%$ in atom N3 and $+2.4 \%$ in atom C8. In all other atoms
the differences are below $2  \%$. For the LUMO state the biggest difference appears in atom
C6, which is $-4.2 \%$, while in the other atoms the differences do not exceed $1.6 \%$.

Regarding cytosine, the HOMO and LUMO energies change slightly (by $+0.1$ eV),
as compared to those of the isolated base. In particular, the predictions of cytosine's
HOMO and LUMO energies within the base-pair conformation are $E_H^C=-8.8$ eV
and $E_L^C=-4.3$ eV, respectively. For the HOMO wavefunction, the differences of
the occupation probabilities in the two base conformations are below $2.5 \%$ for
all atoms. On the contrary, significant changes appear in the LUMO
wavefunction of cytosine within the B-DNA conformation. The occupation
probabilities differ by $+14.6 \%$ in atom C2, $-6.4 \%$
in atoms C4 and C6, $ -4.1 \%$ in atom N3, and $+3.2 \%$ in
atom N1, as compared to the isolated base LUMO, while in the other atoms
the differences are below $2.5 \%$.

For the G-C base-pair, the obtained HOMO energy is $E_H^{GC} = -8.0$ eV.
Several values calculated with methods from first principles have been
reported in the literature: $-7.68$ eV \cite{SS_JACS_1996},
$-7.51$ eV \cite{HC_JACS_1996}, $-7.35$ eV \cite{ZLYY_JCP_2002},
$-7.2$ eV \cite{LCS_JPCA_and_JPCB_01_02}, and $-6.27$ eV \cite{SL_JPCA_2002}.

The calculated LUMO energy here is $E_L^{GC} = -4.5$ eV. Concerning
the first $\pi$-$\pi^*$ transition energy, our prediction is 3.5 eV.
This result is in a reasonable accordance with
the $E_{\pi-\pi^*}$ value obtained in Ref.~\cite{VFMR_JPCB_2006} (4.29 eV).
However, these values differ from the one reported in Ref.~\cite{SL_JPCA_2002}
(6.27 eV), where, similarly to the case of the A-T base-pair,
the HOMO-LUMO gap is attributed to a $\pi$-$\sigma^*$ transition.

From Fig.~\ref{figGC} (bottom) we see that both HOMO and LUMO wavefunctions
of a G-C base-pair are completely localized in guanine.
The complete HOMO localization in guanine results again from the
significantly higher HOMO energy of G in respect to that of C
($E_H^G-E_H^C = 0.8$ eV), as compared with the hopping integral,
$|t_H| = 12$ meV in this case (see Table \ref{table:intrabpt} below).
Similarly, for the LUMO wavefunction of G-C, its complete localization
in guanine is due to the lower LUMO energy of G with respect
to that of C ($E_L^G-E_L^C=-0.2$ eV), as compared with the corresponding
overlap integral of the bases' wavefunctions within the base-pair
($t_L = 16$ meV, see Table~\ref{table:intrabpt}).
These results are in accordance with our findings that $E_H^{GC}=E_H^G$ and
$E_L^{GC}=E_L^G$. They also explain that the electron or hole distribution
within the base-pair is the same with the corresponding distribution in guanine,
as it can be seen by a comparison of the corresponding wavefunctions in the
middle and bottom panels of Fig.~\ref{figGC}. Any small differences in the occupation
probabilities of these plots reflect the changes in guanine's HOMO or LUMO wavefunctions
between the B-DNA and isolated conformations, as it has been discussed above.

It is worth mentioning here that G-C's LUMO wavefunction is the only
HOMO or LUMO of both base-pairs which shows a qualitatively different
behavior when the base distortions within B-DNA are taken into account,
as opposed with the case of the isolated bases.
In particular, considering the structure of the isolated bases and calculating
A-T's HOMO and LUMO or G-C's HOMO, one finds similar results with the
ones presented in this and the previous subsection (for example, A-T's HOMO
is localized in A and LUMO in T, while G-C's HOMO is localized in G).
On the contrary, application of our method for the G-C LUMO, using the structure
of the isolated bases, shows localization mainly in C ($\approx 80 \%$)
but with a non-vanishing amplitude in G ($\approx 20 \%$). In this context
note that the isolated G and C bases exhibit the same LUMO energy, $-4.4$ eV
(see Table~\ref{table:bHL}).
These results seem to be in accordance with the calculations of Ref. \cite{mallajosyula_et_al_2005},
although the nature of the HOMO and LUMO base-pair wavefunctions
(whether it is of $\pi$ character or not) was not mentioned there.

\subsection{Charge transfer hopping parameters}
\label{sec:hopping parameters}

The charge transfer hopping parameters $t_H$ and $t_L$ are obtained below,
as described in Sec.~\ref{sec:theory}, viz.
through the HOMO and LUMO wavefunctions calculated in the previous
subsection and Eqs.~(\ref{tme}) or (\ref{tibp}), for bases or base-pairs,
respectively, with $V_{ij}$ given by Eq.~(\ref{s-k}).

\subsubsection{Description at the base-pair level}
\label{sec:hopping parameters base-pairs}

At first we present the hopping parameters $t_{H/L}^{bp}$ between
successive base-pairs, for the description at the base-pair level
discussed in Sec.~\ref{sec:bpdes} (cf. parameters
$t^{bp (\lambda;\lambda\pm1)}_{H/L}$ in Eq.~(\ref{TBbp})).
The notation $YX$ is used here to denote two successive base-pairs,
according to the following convention for the DNA strands orientation
\begin{eqnarray}
5' & & 3' \nonumber \\
Y & - & Y_{compl} \nonumber \\
X & - & X_{compl} \nonumber  \\
3' & & 5' \; .
\label{bpdimer}
\end{eqnarray}
$X$, $X_{compl}$, $Y$, $Y_{compl}$ denote DNA bases, where $X_{compl}$
and $Y_{compl}$ are the complementary bases of $X$ and $Y$, respectively.
Therefore, the notation $YX$ means that the bases $Y$ and $X$ of
two successive base-pairs ($Y$-$Y_{compl}$ and $X$-$X_{compl}$)
are located at the same strand in the direction $5'-3'$.
$X$-$X_{compl}$ is one base-pair of the base-pair dimer and
$Y$-$Y_{compl}$ is the other base-pair,
separated and twisted by 3.14 {\AA} and $36^{\circ}$, respectively,
relatively to the first base-pair.
For example the notation AC denotes that the base-pair dimer consists of an
adenine-thymine and a cytosine-guanine base-pair, where
one strand contains A and C in the direction $5'-3'$ and the complementary
strand contains T and G in the direction $3'-5'$.

Table \ref{table:interbpTP} summarizes the calculated hopping parameters,
$t_{H/L}^{bp}$, according to our method, for all possible combinations of
successive base-pairs, as well as corresponding values obtained from other
theoretical studies.
Due to the symmetry between base-pair dimers $YX$ and $X_{compl}Y_{compl}$,
the number of different hopping parameters is reduced from sixteen to ten.
In Table \ref{table:interbpTP} base-pair dimers exhibiting the same transfer
parameters are listed together in the first column.
Our predictions for the HOMO, $t_H^{bp}$, and LUMO, $t_L^{bp}$, hopping
parameters are reported in the second and sixth column, respectively.
The third column contains the {\it ab initio} results of Voityuk
{\it et al.}~\cite{VJBR_JCP_2001}
for the magnitudes of HOMO transfer parameters.
Results from Endres {\it et al.}~\cite{endres_et_al_2002}
for a few combinations of successive base-pairs, using the
density-functional-based electronic structure program SIESTA, are presented
for $t_H^{bp}$ and $t_L^{bp}$ in the fourth and seventh columns, respectively.
The same authors show similar results in Fig.~4 of
Ref.~\cite{endres_et_al_2004}. The corresponding values --approximately
extracted from that figure for zero twist angle--
are shown in the fifth and eighth columns. As it has been already mentioned, in
Ref.~\cite{endres_et_al_2004} the transfer parameters have been obtained
using the same theoretical bedrock adopted also in our work --i.e. the
Slater-Koster semi-empirical method and Eqs.~(\ref{s-k}) and (\ref{tibp})--
but with slightly different expressions for $V_{pp\pi}$ and $V_{pp\sigma}$.

\begin{table}[h!]
\caption{Base-pair transfer integrals for all possible combinations of
successive base-pairs, as shown in the first column according to the
notation described in the text, in the direction $5'-3'$.
Hole hopping parameters $t_H^{bp}$ obtained from our calculations (through HOMO wavefunctions),
from Ref.~\cite{VJBR_JCP_2001}, from Ref.~\cite{endres_et_al_2002}, and
from Fig.~4 of Ref.~\cite{endres_et_al_2004},
are presented in the second, third, fourth, and fifth column, respectively.
Electron hopping parameters $t_L^{bp}$ obtained from our calculations (through LUMO wavefunctions),
from Ref.~\cite{endres_et_al_2002}, and from Fig.~4 of Ref.~\cite{endres_et_al_2004},
are shown in the sixth, seventh, and eighth column, respectively.
These quantities represent the parameters
$t^{bp (\lambda;\lambda\pm1)}_{H/L}$ which appear in Eq.~(\ref{TBbp}).
All hopping integrals $t_{H/L}^{bp}$ are given in meV.}
\centerline{
\begin{tabular}{|c|c|c|c|c|c|c|c|} \hline
Base-pair& $t_H^{bp}$ & $|t_H^{bp}|$  &  $t_H^{bp}$
& $t_H^{bp}$ & $t_L^{bp}$ & $t_L^{bp}$ & $t_L^{bp}$ \\
sequence & & \cite{VJBR_JCP_2001} & \cite{endres_et_al_2002} &
\cite{endres_et_al_2004} & & \cite{endres_et_al_2002} & \cite{endres_et_al_2004}  \\ \hline \hline
AA, TT& -8& 26& -70 & $\approx$ -25 & -29&105& $\approx$ 35 \\ \hline
    AT& 20& 50&     &               & 0.5&   & \\ \hline
AG, CT& -5&122&  -71& $\approx$ -50 &   3&112& $\approx$ 35\\ \hline
AC, GT&  2& 27&     &               &  32&   & \\ \hline
    TA& 47& 55&     &               &   2&   & \\ \hline
TG, CA& -4& 26&     &               &  17&   & \\ \hline
TC, GA&-79& 25& -187& $\approx$ -160&  -1& 47& $\approx$ 35\\ \hline
GG, CC&-62& 93& -141& $\approx$ -140&  20& 53& $\approx$ 35\\ \hline
    GC&  1& 78&     &               & -10&   & \\ \hline
    CG&-44& 22&     &               &  -8&   & \\ \hline
\end{tabular} }   \label{table:interbpTP}
\end{table}

We see from Table \ref{table:interbpTP} that according to our calculations (second column),
the higher in magnitude hopping parameters for holes appear in TC, GA, GG, CC, TA,
and CG dimers, where the corresponding transfer matrix elements are around 50 meV
or larger. The lower magnitudes of hole hopping parameters (1-2 meV) occur in
GC, AC,  and GT dimers.
According to Ref.~\cite{VJBR_JCP_2001}, the higher values are found in
AG and CT dimers. In four out of sixteen dimers (GG, CC, CG
and TA) the absolute values of our predictions for $t_H^{bp}$ and those of
Ref.~\cite{VJBR_JCP_2001} differ by no more than a factor of two, while
in other five cases (GC, AC, GT, AG, and CT) our results are
significantly smaller by more than one order of magnitude.
However, it should be noted that more sophisticated theoretical
descriptions~\cite{BV_JPCA_2006} have shown that the approximation used
in Ref.~\cite{VJBR_JCP_2001} in general overestimates the transfer integrals.
Endres {\it et al.} in Refs. \cite{endres_et_al_2002} and
\cite{endres_et_al_2004} provide $t_H^{bp}$ values for a few cases of
successive base-pairs, which are larger in magnitude than
both our results and those of Ref.~\cite{VJBR_JCP_2001},
with the exception of AG and CT where their values are in between
of our results and those of Ref.~\cite{VJBR_JCP_2001}.

Regarding electron hopping parameters (sixth column), we find the highest
magnitudes (around 30 meV) in AC, GT, AA, and TT dimers, while the lowest one
is in AT (less than 1 meV). The most recent $t_L^{bp}$ values of Endres {\it et al.}
in Ref.~\cite{endres_et_al_2004} are closer in magnitude to our predictions,
as compared with their former estimates in Ref.~\cite{endres_et_al_2002}.
For the majority of base-pair dimers, the hole transfer parameters, as calculated
in our work, are larger in magnitude than the corresponding ones for electrons,
indicating probably that hole transport is more favorable than
electron transport along DNA. This is certainly true
for the higher amplitude hopping parameters (larger than 40 meV).

\subsubsection{Description at the single-base level}
\label{sec:hopping parameters bases}

At this level of description a charge carrier is allowed to hop not only between bases
belonging to adjacent base-pairs, but also  from one DNA base to its complementary
base within the same base-pair [see the second term in the right-hand-side
of Eqs. (\ref{TBb})]. In Table~\ref{table:intrabpt} we display these interbase
intra-base-pair hopping parameters. Our calculations show that the magnitudes of hole
and electron transfer integrals within G-C and A-T base-pairs are around 10 meV.
For holes we find equal hopping parameters within G-C or A-T base-pairs, while
for electrons intra-base-pair hopping is more favorable in a G-C base-pair.
First principles results from Ref.~\cite{MA_PRB_2005} show larger magnitudes
of electron hoppings than our predictions (almost four times larger), while for holes
the magnitudes are larger than ours in A-T and smaller in G-C base-pairs
(in both cases the corresponding results differ by a little more than a factor of two).

\begin{table}[h!]
\caption{Intra-base-pair hopping parameters of the two B-DNA base-pairs
(first column), for holes ($t_H^b$, second and third column) and electrons
($t_L^b$, fourth and fifth column). Our predictions for holes (electrons) are
shown in second (fourth) column, while results from Ref. \cite{MA_PRB_2005}
are shown in third (fifth) column, respectively.
These quantities represent the parameters $t^{b(\lambda,1;\lambda,2)}_{H/L}$
and $t^{b(\lambda,2;\lambda,1)}_{H/L}$ which appear in Eq.~(\ref{TBb}).
All hopping parameters $t_{H/L}^b$ are given in meV.}
\centerline{
\begin{tabular}{|c|c|c|c|c|} \hline
Base-pair & $t_H^b$ &$t_H^b$ & $t_L^b$ & $t_L^b$ \\
& & \cite{MA_PRB_2005} &  & \cite{MA_PRB_2005}   \\ \hline \hline
A-T       &     -12 &26 &-9  &34 \\ \hline
G-C       &     -12 &5  & 16 &63 \\ \hline
\end{tabular}  } \label{table:intrabpt}
\end{table}

Now we turn to interbase transfer integrals between bases of adjacent
base-pairs. First we examine intrastrand parameters, i.e. hoppings between
successive bases of the same strand [see the third and fourth terms in the
right-hand-side of Eqs. (\ref{TBb})]. A similar convention in the notation is used
here as previously; $YX$ implies that the DNA bases $Y$ and $X$ are given in
the direction $5'-3'$ of the strand. For example, the notation TG denotes the
strand orientation $5'-$T$-$G$-3'$. The intrastrand interbase hopping parameters
as obtained from our calculations are presented in Table~\ref{table:interbTP}
(in second column for holes and eighth for electrons). Regarding hole transfer,
the higher in magnitude hopping matrix elements (larger than 100 meV) are
found in TT and CT. Other relatively large magnitudes appear
in TC, GC, GA, GT, AT, AC, CC, and GG ($|t_H^b| > 60$ meV).
The lower absolute values of hole hopping parameters correspond to
successive CG, AG, CA, and AA bases ($|t_H^b| < 10$ meV).
Looking at the electron hopping parameters, the highest magnitudes appear
in CT (larger than 60 meV) and CC, GC ($|t_L^b| > 40$ meV),
while the lower ones in AG, AC, AT, and TA ($|t_L^b| < 10$ meV).
In most of the cases electron hopping parameters are smaller in magnitude
than hole transfer parameters between successive bases,
similarly with the case of hopping between successive base-pairs.

Other theoretical predictions for intrastrand transfer integrals have been presented in
Refs. \cite{VRBJ_JPCB_2000,MA_PRB_2005,BV_JPCA_2006,Ratner_JACS_2007,
KWCE_JPCB_2008}, using methods from first principles.
The corresponding values are shown in Table~\ref{table:interbTP}. Because
these methods are usually more reliable for occupied orbitals, there exist many
works computing hole hopping parameters $t_H^b$. In general, our results are in the
range of values obtained by these works.
The {\it ab initio} method used in Ref.~\cite{VRBJ_JPCB_2000}
(similar to that of Ref.~\cite{VJBR_JCP_2001}),
has been shown to overestimate the transfer parameters~\cite{BV_JPCA_2006}.
Indeed, in most of the cases the values provided in Ref.~\cite{VRBJ_JPCB_2000} are larger
than the other $t_H^b$ values of Table~\ref{table:interbTP}.
Regarding electron hopping parameters, first principles results of $t_L^b$ for
a few cases of successive bases have been presented in Ref.~\cite{MA_PRB_2005},
which are in a good agreement with our calculations.

\begin{table}[h!]
\caption{Intrastrand transfer parameters between two successive DNA bases
(first column) in the direction $5'-3'$ of the strand. Hole hopping parameters
$t_H^b$ obtained from our calculations (through HOMO wavefunctions),
and from Refs. \cite{VRBJ_JPCB_2000}, \cite{BV_JPCA_2006},
\cite{Ratner_JACS_2007}, \cite{KWCE_JPCB_2008}, and \cite{MA_PRB_2005},
are presented in the second, third, fourth, fifth, sixth, and seventh column, respectively.
Electron hopping parameters $t_L^b$ obtained from our calculations (through LUMO
wavefunctions) and from Ref.~\cite{MA_PRB_2005}, are shown in the eighth and
ninth column, respectively. These quantities represent the parameters
$t^{b (\lambda,i;\lambda\pm1,i)}_{H/L}$ which appear in Eq.~(\ref{TBb}).
All hopping parameters $t_{H/L}^b$ are given in meV.}
\centerline{
\begin{tabular}{|c|c|c|c|c|c|c|c|c|} \hline
 Base &$t_H^b$ &$|t_H^b|$ &$|t_H^b|$ &$t_H^b$ & $|t_H^b|$ &
$t_H^b$ &$t_L^b$ &$t_L^b$ \\
sequence & & \cite{VRBJ_JPCB_2000} & \cite{BV_JPCA_2006} & \cite{Ratner_JACS_2007}
& \cite{KWCE_JPCB_2008} & \cite{MA_PRB_2005} & & \cite{MA_PRB_2005} \\ \hline \hline
    AA &-8   &30   &4  &-4   &8  &21   & 16  &25\\ \hline
    AT &68   &105  &   &-63  &28 &     &  7  &      \\ \hline
    AG &-5   &49   &44 &-10  &37 &     &  1  &      \\ \hline
    AC &68   &61   &   &42  &30  &     & -3  &      \\ \hline
    TA &26   &86   &   &-31  &64 &     & -7  &      \\ \hline
    TT &-117 &158  &   & 72  &93 &-98  & -30 &-23\\ \hline
    TG &28   &85   &61 & 18  &70 &     & -17 &      \\ \hline
    TC &-86  &76   &   &-28  &52 &     &  22 &      \\ \hline
    GA &-79  &89   &36 &-77  &52 &     &  30 &      \\ \hline
    GT &73   & 137   &81 & 141 &49 &     & -32 &      \\ \hline
    GG &-62  &84   &51 & 53  &61 &-114 &  20 &20    \\ \hline
    GC &80   &110  &   &-114 &57 &     &  43 &      \\ \hline
    CA &5    &29   &   &-2   &5  &     & -12 &      \\ \hline
    CT &-107 &100  &   &-55  &33 &     &  63 &      \\ \hline
    CG &-1   &42   &   & 9   &31 &     &  15 &      \\ \hline
    CC &-66  &41   &   & 22  &26 &-21  & -47 &-60\\ \hline
\end{tabular}  } \label{table:interbTP}
\end{table}

Finally we present interstrand transfer parameters describing "diagonal" interbase
hoppings between diagonally located bases of adjacent base-pairs which belong to
opposite strands [see the fifth and sixth terms in the right-hand-side of Eqs. (\ref{TBb})].
Here it is necessary to distinguish two sets of diagonal interstrand hoppings, depending on
the strand orientation. For the conformation of two successive base-pairs shown
in (\ref{bpdimer}), the one set of diagonal interstrand transfer parameters refers to
hoppings in the direction $3'-3'$ (i.e. the path $3'-Y_{compl} \cdots X-3'$) and the other
set refers to hoppings in the direction $5'-5'$ (i.e. the path $5'-Y \cdots X_{compl}-5'$).

\begin{table}[h!]
\caption{$3'-3'$ interstrand hopping parameters between two diagonally located DNA
bases of adjacent base-pairs (first column). Hole hopping parameters $t_H^{b}$
obtained from our calculations (through HOMO wavefunctions) and from
Refs. \cite{Ratner_JACS_2007}, \cite{MA_PRB_2005}, are presented in the second,
third, and fourth column, respectively. Electron hopping parameters $t_L^{b}$ obtained
from our calculations (through LUMO wavefunctions) and from Ref.~\cite{MA_PRB_2005},
are shown in the fifth and sixth column, respectively. These quantities represent the
parameters $t^{b (\lambda,i;\lambda\pm1,j)}_{H/L}$ ($i \neq j$) which appear in Eq.~(\ref{TBb}).
All hopping parameters $t_{H/L}^b$ are given in meV.}
\centerline{
\begin{tabular}{|c|c|c|c|c|c|} \hline
Base &$t_H^b$ &$t_H^b$ &$|t_H^b|$ &$t_L^b$ &$t_L^b$  \\
sequence & & \cite{Ratner_JACS_2007} & \cite{MA_PRB_2005} & & \cite{MA_PRB_2005}
 \\  \hline  \hline
AA    & 48& 49&     & 29 &     \\ \hline
AT, TA&-3 &-7 & 9   & 3  & -13 \\ \hline
AG, GA&-3 &-11&     &-6  &     \\ \hline
AC, CA&-5 & 17&     &-3  &     \\ \hline
TT    &0.5&  6&     &0.2 &     \\ \hline
TG, GT& 5 &-14&     & 2  &     \\ \hline
TC, CT&0.5&  4&     &-0.2&     \\ \hline
GG    &-44&-32&     &-5  &     \\ \hline
GC, CG& 4 & 22&48   &-4  & -15 \\ \hline
CC    & 1 & 10&     &0.3 &     \\ \hline
\end{tabular} }   \label{table:interbD3TP}
\end{table}

Table~\ref{table:interbD3TP} shows results for interstrand transfer integrals in the direction
$3'-3'$ for all possible combinations of bases, while Table~\ref{table:interbD5TP} presents
corresponding results in the direction $5'-5'$. Our calculations yield relatively small
magnitudes for these matrix elements. In general the obtained values are of the order
of meV or tenths of meV, for both electrons and holes, apart from the cases of
$3'-A \cdots A-3'$ and $3'-G \cdots G-3'$ for holes (exhibiting $|t_H^b|$ between
40-50 meV) and $3'-A \cdots A-3'$ for electrons (with $t_L^b$ around 30 meV).
Theoretical predictions for hole interstrand hopping parameters $t_H^b$ have been
also presented for all combinations of bases in Ref.~\cite{Ratner_JACS_2007}.
Other results from first principles methods
obtained for a few cases in Refs. \cite{VRBJ_JPCB_2000} (for holes) and  \cite{MA_PRB_2005} (both for electrons and holes),
show usually larger magnitudes in comparison to our calculations.

\begin{table}[h!]
\caption{$5'-5'$ interstrand hopping parameters between two diagonally located DNA
bases of adjacent base-pairs (first column). Hole hopping parameters $t_H^{b}$
obtained from our calculations (through HOMO wavefunctions) and from
Refs. \cite{Ratner_JACS_2007}, \cite{VRBJ_JPCB_2000}, \cite{MA_PRB_2005},
are presented in the second, third, fourth, and fifth column, respectively. Electron hopping
parameters $t_L^{b}$ obtained from our calculations (through LUMO wavefunctions)
and from Ref.~\cite{MA_PRB_2005} are shown in the sixth and seventh column,
respectively. These quantities represent the parameters
$t^{b (\lambda,i;\lambda\pm1,j)}_{H/L}$ ($i \neq j$) which appear in Eq.~(\ref{TBb}).
All hopping parameters $t_{H/L}^b$ are given in meV.}
\centerline{
\begin{tabular}{|c|c|c|c|c|c|c|} \hline
Base &$t_H^b$&$t_H^b$ &
$|t_H^b|$ &$t_H^b$ &$t_L^b$ &$t_L^b$  \\
sequence& & \cite{Ratner_JACS_2007} & \cite{VRBJ_JPCB_2000} & \cite{MA_PRB_2005} &  &
 \cite{MA_PRB_2005}  \\ \hline \hline
AA    & 2 & 31& 31-35  &      &  6 &    \\ \hline
AT, TA& 9 & 7 & 16-20  & -11  &  2 &-10 \\ \hline
AG, GA& 4 &-13& 19-24  &      &  3 &    \\ \hline
AC, CA& 5 &-1 &     &      & -2 &    \\ \hline
TT    & 4 & 1 &  3-4  &      &  2 &    \\ \hline
TG, GT& 5 &-9 &     &      &  3 &    \\ \hline
TC, CT& 2 &0.3&     &      & -2 &    \\ \hline
GG    & 3 & 12& 19  &      & -2 &    \\ \hline
GC, CG& 4 &  2&     & -8   & -3 &-12 \\ \hline
CC    & 1 &  1& 1   &      &  2 &    \\ \hline
\end{tabular} }   \label{table:interbD5TP}
\end{table}

\section{Conclusions}
\label{sec:conclusions}

We have systematically studied all the tight-binding parameters
which are necessary for the description of charge transfer along DNA.
The $\pi$ electronic structure of the four DNA bases
(adenine, thymine, cytosine, and guanine) has been calculated by using
the LCAO method, employed with a novel parametrization.
In addition, we have presented the HOMO and LUMO of
the two B-DNA base-pairs (adenine-thymine and guanine-cytosine),
using a similar process of linear combination of molecular orbitals.
Taking into account the slight deformation of bases
within the base-pairs of B-DNA (compared to the isolated bases)
we find that
(a) for the A-T base-pair the HOMO resides in A and the LUMO in T, while
(b) for the G-C base-pair both HOMO and LUMO reside in G.

Our theoretical approach predicts the $\pi$ HOMO energy of DNA bases
with a deviation smaller than 0.3 eV in comparison with the experimental
values. In particular, for the bases of guanine and cytosine our results coincide
with the experimental ones, while for thymine and
adenine the difference ranges from 0-0.2 eV and 0.2-0.3 eV, respectively.
Regarding the first $\pi$-$\pi^*$ transition energy the deviations are larger:
0-0.2 eV for cytosine, 0.4-0.5 eV for thymine, 0.5-0.7 eV for guanine, and
0.7-1.0 eV for adenine. Our results for the HOMO and LUMO energies and
wavefunctions of bases and also of base-pairs have been compared with other
theoretical calculations, using methods from first principles.

We provide estimates for the complete set of hopping parameters (both for
electrons and holes) between successive base-pairs and between neighboring bases
(i.e., successive bases in the same strand, complementary bases within a base-pair,
and adjacent base-pairs' diagonally located bases in opposite strands) in B-DNA,
including all possible combinations of them. Our predictions for the transfer parameters
are compared with other theoretical estimates, when available, and in most
cases there is an agreement in the order of magnitude of the results.
In general, the hopping parameters for hole transfer obtained through the HOMO
wavefunctions, are higher in magnitude compared to the ones for electron transfer
obtained through the LUMO wavefunctions. These theoretical calculations show
that probably  hole transport along DNA is more favorable than electron transport.

The microscopic quantities calculated in this work, i.e. the HOMO and LUMO energies
as well as the hopping matrix elements, provide all necessary parameters for a
tight-binding phenomenological description of charge transfer along the DNA
double helix, based either on bases' or base-pairs' $\pi$ molecular overlap.
Taking advantage of such a description, at a mesoscopic level,
the temporal and spatial evolution of electron or hole transport along DNA
can be examined in experimentally relevant time and length scales.
Furthermore, important measurable quantities like hole/electron transmission
coefficients and conductivities can be calculated for any DNA segment
(of whatever base sequence) under consideration, since there are well-established
techniques to compute such a quantities from tight-binding models.

{\it Acknowledgements} We acknowledge useful discussions with N. Lathiotakis
and support from the C. Carath\'{e}odory program C155 of University of Patras.


\begin{thebibliography}{39}

\bibitem{dekker_ratner_2001} C.~Dekker and M.~Ratner,
Phys. World, August 2001, 29 (2001).

\bibitem{keren_et_al_2002} K.~Keren, M.~Krueger, R.~Gilad, G.~Ben-Yoseph,
U.~Sivan, and E.~Braun, Science {\bf 297}, 72 (2002).

\bibitem{seeman_2003} N.C.~Seeman, Nature {\bf 421}, 427 (2003).

\bibitem{endres_et_al_2004} R.G.~Endres, D.L.~Cox, and R.R.P.~Singh,
Rev. Mod. Phys. {\bf 76}, 195 (2004).

\bibitem{burrows_muller_1998} C.J.~Burrows and J.G.~Muller,
Chem. Rev. {\bf 98}, 1109 (1998).

\bibitem{cadet_et_al_1994} J. Cadet in {\it DNA Adducts: Identification
and Significance}, eds. K. Hemminki, A. Dipple, D.E.F. Shiker, F.F. Kadlubar,
D. Segerback, and H. Bartsch, IARC, Lyon (1994).

\bibitem{dandliker_et_al_1997} P.J. Dandliker, R.E. Holmlin, J.K. Barton,
Science {\bf 275}, 1465 (1997).

\bibitem{rajski_et_al_2000} S.R.~Rajski, B.A.~Jackson, and J.K.~Barton,
Mutation Research {\bf 447}, 49 (2000).

\bibitem{zhang_et_al_2002} Y.~Zhang, R.H.~Austin, J.~Kraeft, E.C.~Cox, and
N.P.~Ong, Phys. Rev. Lett. {\bf 89}, 198102 (2002).

\bibitem{storm_et_al_2001} A.J.~Storm, J.~van~Noort, S.~de~Vries, and
C.~Dekker, Appl. Phys. Lett. {\bf 79}, 3881 (2001).

\bibitem{de_pablo_et_al_2000} P.J.~de~Pablo, F.~Moreno-Herrero, J.~Colchero,
J.~G\'{o}mez Herrero, P.~Herrero, A.M.~Bar\'{o}, P.~Ordej\'{o}n, J.M.~Soler,
and E.~Artacho, Phys. Rev. Lett. {\bf 85}, 4992 (2000).

\bibitem{yoo_et_al_2001} K.-H.~Yoo, D.H.~Ha, J.-O.~Lee, J.W.~Park, Jinhee~Kim,
J.J.~Kim, H.-Y.~Lee, T.~Kawai, and Han~Yong ~Choi,
Phys. Rev. Lett. {\bf 87}, 198102 (2001).

\bibitem{cai_et_al_2000} L.~Cai, H.~Tabata, and T.~Kawai,
Appl. Phys. Lett. {\bf 77}, 3105 (2000).

\bibitem{fink_schonenberger_1999} H.W.~Fink and C.~Schonenberger,
Nature {\bf 398}, 407 (1999).

\bibitem{tran_et_al_2000} P.~Tran, B.~Alavi, and G.~Gruner,
Phys. Rev. Lett. {\bf 85}, 1564 (2000).

\bibitem{kutnjak_et_al_2003} Z.~Kutnjak, C.~Filipi\v{c}, R.~Podgornik,
L.~Nordenski\"{o}ld, and N. Korolev, Phys. Rev. Lett. {\bf 90}, 098101 (2003).

\bibitem{macnaughton_et_al_2006} J.B.~MacNaughton, A.~Moewes, J.S.~Lee,
S.D.~Wettig, H.-B.~Kraatz, L.Z.~Ouyang, W.Y.~Ching, and E.Z.~Kurmaev,
J. Phys. Chem. B {\bf 110}, 15742 (2006).

\bibitem{mallajosyula_et_al_2005} S.S.~Mallajosyula, A.~Datta, and
S.K.~Pati, Synth. Met. {\bf 155}, 398 (2005).

\bibitem{MBKEF_PRB_2002} P. Maragakis, R.L. Barnett, E. Kaxiras, M. Elstner,
and T. Frauenheim, Phys. Rev. B {\bf 66}, 241104 (2002).

\bibitem{rakitin_et_al_2001} A.~Rakitin, P.~Aich, C.~Papadopoulos,
Yu.~Kobzar, A.S.~Vedeneev, J.S. Lee, and J.M. Xu,
Phys. Rev. Lett. {\bf 86}, 3670 (2001).

\bibitem{fuentes-cabrera_et_al}
M. Fuentes-Cabrera, B.G. Sumpter, and J.C. Wells,
J. Phys. Chem. B {\bf 109}, 21135 (2005);
M. Fuentes-Cabrera, B.G. Sumpter, P. Lipkowski, and J.C. Wells,
J. Phys. Chem. B {\bf 110}, 6379 (2006);
M. Fuentes-Cabrera, X. Zhao, P.C. Kent, and B.G. Sumpter,
J. Phys. Chem. B {\bf 111}, 9057 (2007).

\bibitem{Barton1996ab} M.R.~Arkin, E.D.A.~Stemp, R.E.~Holmlin, J.K.~Barton,
A.~H\"{o}rmann, E.J.C.~Olson, and P.F.~Barbara, Science {\bf 273}, 475 (1996);
D.B.~Hall, R.E.~Holmlin, and J.K.~Barton, Nature {\bf 382}, 731 (1996).

\bibitem{porath_et_al_2000} D. Porath, A. Bezryadin, S. de Vries,
and C. Dekker, Nature {\bf 403}, 635 (2000).

\bibitem{adesi_et_al}
Ch. Adesi, S. Walch, and M.P. Anantram, Phys. Rev. B {\bf 67}, 081405 (R) (2003);
Ch. Adesi and M.P. Anantram, Appl. Phys. Lett. {\bf 82}, 2353 (2003).

\bibitem{RC} S.V.~Rakhmanova and E.M.~Conwell,
J. Phys. Chem. B {\bf 105}, 2056 (2001).

\bibitem{zhang_et_al_2_2002} W. Zhang, A.O. Govorov, and S.E. Ulloa,
Phys. Rev. B {\bf 66}, 060303 (2002).

\bibitem{KKB} S. Komineas, G. Kalosakas, and A.R. Bishop,
Phys. Rev. E {\bf 65}, 061905 (2002);
P.~Maniadis, G.~Kalosakas, K.\O.~Rasmussen,
and A.R.~Bishop, Phys. Rev. B {\bf 68}, 174304 (2003).

\bibitem{singh_2004} M.R.~Singh, J. Biomater. Sci. Polymer Edn. {\bf 15}, 1533 (2004).

\bibitem{triberis+} G.P.~Triberis, C.~Simserides, and V.C.~Karavolas,
J. Phys.: Condens. Matter {\bf 17}, 2681 (2005);
G.P.~Triberis and M.~Dimakogianni, J. Phys.: Condens. Matter {\bf 21}, 035114 (2009).

\bibitem{ye_et_al_1999} Y.-J.~Ye, R.-S.~Chen, A.~Martinez, P.~Otto, J.~Ladik,
Solid State Commun. {\bf 112}, 139 (1999).

\bibitem{schuster} P.T. Henderson, D. Jones, G. Hampikian, Y. Kan,
and G.B. Schuster, Proc. Natl. Acad. Sci. USA {\bf 96}, 8353 (1999).

\bibitem{bruinsma_et_al_2000} R.~Bruinsma, G.~Gr\"{u}ner, M.R.~D'Orsogna,
and J.~Rudnick, Phys. Rev. Lett. {\bf 85}, 4393 (2000).

\bibitem{KRB} G.~Kalosakas, K.\O.~Rasmussen, and A.R. Bishop,
J. Chem. Phys. {\bf 118}, 3731 (2003);
Synth. Met. {\bf 141}, 93 (2004).

\bibitem{yu_song_2001} Z.G.~Yu and X.~Song, Phys. Rev. Lett. {\bf 86}, 6018 (2001).

\bibitem{hennig+} D.~Hennig, Eur. Phys. J. B {\bf 30}, 211 (2002);
D. Hennig, E.B. Starikov, J.F.R. Archilla, and F. Palmero,
J. Biol. Phys. {\bf 30}, 227 (2004).

\bibitem{KNF+MKRB} G.~Kalosakas, K.L.~Ngai, and S.~Flach,
Phys. Rev. E {\bf 71}, 061901 (2005);
P.~Maniadis, G.~Kalosakas, K.\O.~Rasmussen,
and A.R.~Bishop, Phys. Rev. E {\bf 72}, 021912 (2005).

\bibitem{lima}  E. Diaz, R.P.A. Lima, and F. Dominguez-Adame,
Phys. Rev. B {\bf 78}, 134303 (2008).

\bibitem{HKS} L.G.D. Hawke, G. Kalosakas, and C. Simserides,
Mol. Phys. {\bf 107}, 1755 (2009).

\bibitem{flavin} L.G.D. Hawke, C. Simserides, and G. Kalosakas,
Mater. Sci. Eng. B, in press (2009), doi:10.1016/j.mseb.2009.02.012

\bibitem{harrison} W.A.~Harrison, {\it Electronic structure and the
properties of solids}, 2nd edition, Dover, New York (1989);
{\it Elementary electronic structure}, World Scientific (1999).

\bibitem{slater_koster_1954} J.C.~Slater and G.F.~Koster,
Phys. Rev. {\bf 94}, 1498 (1954).

\bibitem{MA} M. Menon and R.E. Allen, Phys. Rev. B {\bf 38}, 6196 (1988).

\bibitem{LA} N.~Lathiotakis and A.N.~Andriotis,
Solid State Comm. {\bf 87}, 871 (1993);
M. Menon, J. Connolly, N. Lathiotakis, and A. Andriotis,
Phys. Rev. B {\bf 50}, 8903 (1994).

\bibitem{gstruct} http://chemistry.gsu.edu/Glactone/PDB/pdb.html

\bibitem{HC Chem. Phys. Let 1975} N.S. Hush and A.S. Cheung,
Chem. Phys. Lett. {\bf 34}, 11 (1975).


\bibitem{Tetrahedron 45_1975}
 G. Lauer, W. Schafer, and A. Schweig, Tetrahedron Lett. {\bf 45}, 3939 (1975).

\bibitem{DWMG}  D. Dougherty, K. Wittel, J. Meeks, and S.P. McGlynn,
J. Am. Chem. Soc. {\bf 98}, 3815 (1976).

\bibitem{JACS 10_1980 and JPC 84_1980}
J. Lin, C. Yu, S. Peng, I. Akiyama, K. Li, L.K. Lee, P.R. LeBreton,
J. Am. Chem. Soc. {\bf 102}, 4627 (1980);
J. Lin, C. Yu, S. Peng, I. Akiyama, K. Li, L.K. Lee, P.R. LeBreton, J. Phys. Chem. {\bf 84}, 1006 (1980).

\bibitem{JMolSt 214_1989} S. Urano, X. Yang, and P. R. LeBreton,
Journal of Molecular Sructure {\bf 214}, 315 (1989).

\bibitem{VGCD_Biopolymers_1963}
D. Voet, W.B. Gratzer, R.A. Cox, and P. Doty, Biopolymers {\bf 1}, 193 (1963).

\bibitem{CPT_JPC_1965}
L.B. Clark, G.G. Peschel, and I. Tinoco Jr., J. Phys. Chem. {\bf 69}, 3615 (1965).

\bibitem{MRE_PNAS_67} D.W. Miles, R.K. Robins, and H. Eyring,
Proc. Natl. Acad. Sci. USA {\bf 57}, 1139 (1967).

\bibitem{YF_Biopolymers_1968}
T. Yamada and H. Fukutome, Biopolymers {\bf 6}, 43 (1968).

\bibitem{VRBD_JACS_1968}  W. Voelter, R. Records, E. Bunnenberg,
and C. Djerassi, J. Am. Chem. Soc. {\bf 90}, 6163 (1968).

\bibitem{FF_JACS_1971}
A.F. Fucaloro and L.S. Forster, J. Am. Chem. Soc. {\bf 93}, 6443 (1971).

\bibitem{BM_BIOPOLYMERS_75}
W.C. Brunner and M.F. Maestre, Biopolymers {\bf 14}, 555 (1975).

\bibitem{SJ_Biopolymers_1977}
C.A. Sprecher and W. C. Johnson, Biopolymers {\bf 16}, 2243 (1977).

\bibitem{RFFK_Biopolymers_1978}
K. Raksanyi, I. Foldvary, J. Fidy, and L. Kittler, Biopolymers {\bf 17}, 887 (1978).

\bibitem{MN}
Y. Matsuoka and B. Norden, J. Phys. Chem. {\bf 86}, 1378 (1982).

\bibitem{ZNC_JACS_1985}
F. Zaloudek, J.S. Novros, and L.B. Clark, J. Am. Chem. Soc. {\bf 107}, 7344 (1985).

\bibitem{Clark_JPC_1990} L.B. Clark, J. Phys. Chem. {\bf 94}, 2873 (1990).

\bibitem{HBAN_JACS_97} A. Holmen, A. Broo, B. Albinsson, and B. Norden,
J. Am. Chem. Soc. {\bf 119}, 12240 (1997).

\bibitem{CBCS_JPC_1992} A.O. Colson, B. Besler, D.M. Close,
and M.D. Sevilla, J. Phys. Chem. {\bf 96}, 661 (1992).

\bibitem{SS_JACS_1996} H. Sugiyama and I. Saito, J. Am. Chem. Soc. {\bf 118}, 7063 (1996).

\bibitem{HC_JACS_1996} M. Hutter and T. Clark,
J. Am. Chem. Soc. {\bf 118}, 7574 (1996).

\bibitem{FR_JACS_1995} M.P. F\"{u}lscher and B.O. Roos,
J. Am. Chem. Soc. {\bf 117}, 2089 (1995).

\bibitem{LFR_JACS_1995} J. Lorentzon, M.P. F\"{u}lscher, and B.O. Roos,
J. Am. Chem. Soc. {\bf 117}, 9265 (1995).

\bibitem{FSR_JACS_1997} M.P. F\"{u}lscher, L. Serrano-Andres, and B.O. Roos,
J. Am. Chem. Soc. {\bf 119}, 6168 (1997).

\bibitem{SM_CP_1999} M.K. Shukla and P.C. Mishra, Chem. Phys. {\bf 240}, 319 (1999).

\bibitem{MTT_JPCA_2001} B. Mennucci, A. Toniolo, and J. Tomasi,
J. Phys. Chem. A. {\bf 105}, 4749 (2001).

\bibitem{LCS_JPCA_and_JPCB_01_02}
X. Li, Z. Cai, and M.D. Sevilla, J. Phys. Chem. B {\bf 105}, 10115 (2001);
X. Li, Z. Cai, and M.D. Sevilla, J. Phys. Chem. A {\bf 106}, 9345 (2002).

\bibitem{SD_EPJD_2002} A.L. Sobolewski and W. Domcke,
Eur. Phys. J. D {\bf 20}, 369 (2002).

\bibitem{IBOKR_JACS_2002} N. Ismail, L. Blancafort, M. Olivucci,
B. Kohler, and M.A. Robb, J. Am. Chem. Soc. {\bf 124}, 6818 (2002).

\bibitem{SL_JPHA_2002} M.K. Shukla and J. Leszczynski,
J. Phys. Chem. A. {\bf 106}, 11338 (2002).

\bibitem{SL_JCC_2004} M.K. Shukla and J. Leszczynski,
Journal of Computational Chemistry {\bf 25}, 768 (2004).

\bibitem{VFMR_JPCB_2006} D. Varsano, R. Di Felice, M.A.L. Marques, and
A. Rubio, J. Phys. Chem. B {\bf 110}, 7129 (2006).

\bibitem{FKH_JPCA_2007} T. Fleig, S. Knecht, and C. H\"{a}ttig,
J. Phys. Chem. A {\bf 111}, 5482 (2007).

\bibitem{ZLYY_JCP_2002} H. Zhang, X.Q. Li, P. Ham, X.Y. Yu, and Y.J. Yan,
J. Chem. Phys. {\bf 117}, 4578 (2002).

\bibitem{SL_JPCA_2002} M.K. Shukla and J. Leszczynski,
J. Phys. Chem. A {\bf 106}, 4709 (2002).

\bibitem{Jmol} http://www.jmol.org/

\bibitem{CKK_JPCB} T. Cramer, S. Krapf, and T. Koslowski,
J. Phys. Chem. B {\bf 108}, 11812 (2004).

\bibitem{FCM_PRB} R. Di Felice, A. Calzolari, and E. Molinari,
Phys. Rev. B {\bf 65}, 045104 (2001).

\bibitem{VJBR_JCP_2001} A.A. Voityuk, J. Jortner, M. Bixon, and N. R\"{o}sch,
J. Chem. Phys. {\bf 114}, 5614 (2001).

\bibitem{endres_et_al_2002} R.G.~Endres, D.L.~Cox, and R.R.P.~Singh,
arXiv:cond-mat/0201404v2 (2002).

\bibitem{BV_JPCA_2006} L. Blancafort and A.A. Voityuk, J. Phys. Chem. A
{\bf 110}, 6426 (2006).

\bibitem{MA_PRB_2005} H. Mehrez and M.P. Anantram, Phys. Rev. B
{\bf 71}, 115405 (2005).

\bibitem{VRBJ_JPCB_2000} A.A. Voityuk, N. R\"{o}sch, M. Bixon, and J. Jortner,
J. Phys. Chem. B {\bf 104}, 9740 (2000).

\bibitem{Ratner_JACS_2007} K. Senthilkumar, F.C. Grozema, C.F. Guerra,
F.M. Bickelhaupt, F.D. Lewis, Y.A. Berlin, M.A. Ratner, and L.D. A. Siebbeles,
J. Am. Chem. Soc. {\bf 127}, 14894 (2005).

\bibitem{KWCE_JPCB_2008} T. Kubar, P.B. Woiczikowski, G. Cuniberti, and M. Elstner,
J. Phys. Chem. B {\bf 112}, 7937 (2008).

\end{thebibliography}
\end{document}